\documentclass[]{aa} 
\usepackage{graphics,graphicx}
\usepackage{natbib}
\usepackage{txfonts}
\usepackage{xcolor}
\usepackage{comment}
\usepackage[usestackEOL]{stackengine} 
\usepackage{hyperref} 
\newcommand{\bl}[1]{\mbox{\boldmath$ #1 $}}

\begin{document}


\title{Computing the gravitational potential on nested meshes using the convolution method\thanks{Code can be viewed at \href{https://github.com/jamesmckevitt/CM4NG}{https://github.com/jamesmckevitt/CM4NG}.}}

\author{Eduard I. Vorobyov\inst{1}, James McKevitt\inst{1}, Igor Kulikov\inst{2}, Vardan Elbakyan\inst{3,4}}

\institute{University of Vienna, Department of Astrophysics, T\"urkenschanzstrasse 17, Vienna, 1180, Austria;
\email{eduard.vorobiev@univie.ac.at} 
\and
Institute of Computational Mathematics and Mathematical Geophysics SB RAS, Lavrentieva ave., 6, Novosibirsk, 630090 Russia
\and
Research Institute of Physics, Southern Federal University, Stachki Ave. 194, Rostov-on-Don 344090, Russia  
\and
School of Physics, University of Leicester, Leicester, LE1 7RH, UK }

\date{}

   
   \titlerunning{Gravitational potential on nested meshes}
   \authorrunning{Vorobyov et al.}

\abstract
{}
{Our aim is to derive a fast and accurate method for computing the gravitational potential of astrophysical objects with high contrasts in density, for which nested or adaptive meshes are required.}
{We present an extension of the convolution method for computing the gravitational potential to the nested Cartesian grids. The method makes use of the convolution theorem to compute the gravitational potential using its integral form.}
{A comparison of our method with the iterative outside-in conjugate gradient and generalized minimal residual methods for solving the Poisson equation using nonspherically symmetric density configurations has shown a comparable performance in terms of the errors relative to the analytic solutions. However, the convolution method is characterized by several advantages and outperforms the considered iterative methods by factors 10--200 in terms of the runtime, especially when graphics processor units are utilized. The convolution method also shows an overall second-order convergence, except for the errors at the grid interfaces where the convergence is linear.}
{High computational speed and ease in implementation can make the convolution method a preferred choice when using a large number of nested grids. The convolution method, however, becomes more computationally costly if the dipole moments of tightly spaced gravitating objects are to be considered at coarser grids.} 

\maketitle

\section{Introduction}

Gravity, together with magnetic fields and radiation, is a key mechanism that determines the formation and evolution of many astrophysical objects from planets to clusters of galaxies. On rare occasions the gravity force can be found analytically \citep{BT1987}. In practice, various numerical methods are often employed to compute the gravitational potential using the Poisson equation \citep[e.g.,][]{BT1987,Press1992,Bodenheimer2007}. The gravity force can then be calculated from the gradient of the potential using discretization schemes with various levels of complexity \cite[e.g.,][]{StoneNorman1992,KulikovVorobyov2016,Wang2020}. The required computational resources are often substantial and may take up a large portion of the entire runtime of a three-dimensional numerical hydrodynamics code. Therefore, fast methods such as the fast Fourier transform (FFT) are often a method of choice if the geometry of the system permits regular meshes with equidistant cells \citep{Press1992}.  

In many astrophysical applications, the FFT method is difficult to apply because of the requirement for irregular computational meshes that can better resolve the spatial regions where matter accumulates and density increases. For example, the gravitational collapse of interstellar clouds involves an increase in the gas density in the cloud interiors by many orders of magnitude \citep[e.g.,][]{Masunaga1998,Larson2003}.  Extensions of the FFT algorithm to nonequidistant meshes exist \citep{Potts2004} but they are usually much slower than their classical counterparts. In such situations, nested meshes or, more generally, adaptive meshes are often used to achieve a better numerical resolution in the innermost regions \citep[e.g.,][]{Matsumoto2003,Machida2005,Tomida2015, Hennebelle2018,Commercon2020}. 

Finding the gravitational force on nested meshes is a complicated task. A contemporary approach involves solving for the Poisson equation first on the coarsest grid. The result is then used to calculate the gravitational potential on the refined grid using, for the boundary conditions, the potential at the coarsest grid \citep[e.g.,][]{Matsumoto2003,Guillet2011,Wang2020}. Multigrid methods are often used on each level to improve the accuracy, especially when tightly spaced gravitating objects are present on the deepest levels of refinement \citep{Matsumoto2003,2007Matsumoto,1998Truelove,2020Stone}.
Moving in this fashion from the coarsest to the most refined grid, the gravitational potential on the entire nested mesh can be retrieved. However, this ``outside-in'' approach has a potentially serious drawback -- this algorithm is difficult to parallelize because computations on a refined grid cannot be done until the potential on a coarse grid is found. 
According to our experience, calculating the gravity force in numerical hydrodynamics equations on nested meshes with a moderate base resolution $N\le64^3$ may take a large fraction of the entire computational time. Therefore, developing alternative methods is worthwhile.



Here, we present a conceptually different method of computing the gravitational force on nested meshes, which employs the integral form of the gravitational potential and the convolution theorem to solve the resulting triple sum. 
To the best of our knowledge, this method has not been explicitly described in the literature (but see \citet{1990Ruffert}), although its application to regular (non-nested) meshes in astrophysics is well known \citep[e.g.,][]{MullerKley2012,Vorobyov2020,Hahn2020}. The layout of this work is as follows. In Sect.~\ref{inside-out} we briefly outline the outside-in solution procedure using the conjugate gradient method and provide the test results on two nonspherically symmetric mass distributions. In Sect.~\ref{convolution} we review the general convolution method and in Sect.~\ref{nested} we describe its extension to nested meshes. The comparison of models is provided in Sect.~\ref{Compare}. The analysis of runtimes is given in Sect.~\ref{Sect:performance}. Our main conclusions are summarized in Sect.~\ref{conclude}.

\section{Outside-in conjugate gradient solution procedure}
\label{inside-out}
\begin{figure}[h]
\centering
\includegraphics[scale=0.45]{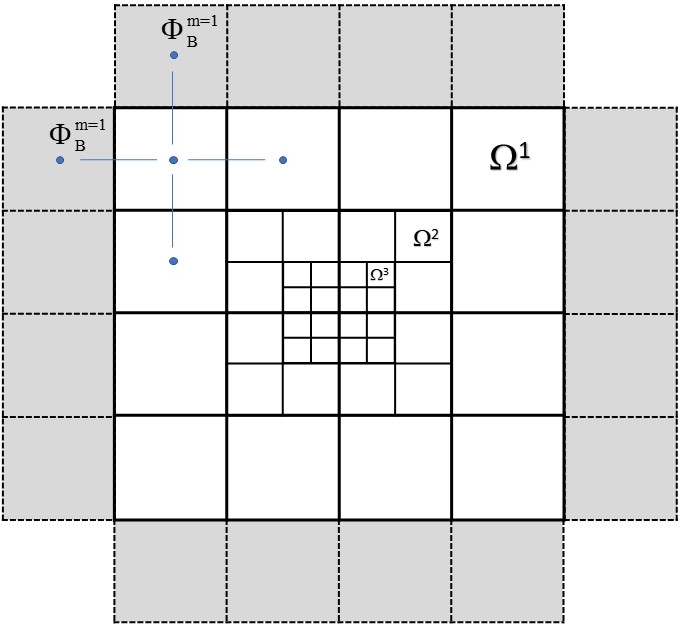}
\caption{Example of a nested mesh with three levels of refinement: $\Omega^1$, $\Omega^2$, and $\Omega^3$. The gray-shaded cells represent boundary layers for the coarsest grid; the gravitational potential at these boundary layers has to be found { a priori}.
}
\label{fig:1}
\end{figure}

In this section, we briefly outline the outside-in solution method for the gravitational potential $\Phi$ on nested meshes. This solution procedure is used merely for comparison with an alternative method that we propose later in Sect.~\ref{convolution}. In general, the gravitational potential of an arbitrary mass distribution can be found by solving for the Poisson equation
\begin{equation}
\label{poisson:eq}
    \nabla \cdot \nabla \Phi = \triangle \Phi = 4 \pi G \rho,
\end{equation}
where $\triangle$ is the Laplace operator, $G$ is the gravitational constant, and $\rho$ is the mass volume density. Figure \ref{fig:1} presents a simple example of three nested meshes on the Cartesian two-dimensional grid $(x,y)$. For simplicity we show only two dimensions. However, a third dimension will be added in practical tests.
By default, the outer coarsest grid $\Omega^1$ encompasses the entire computational domain with a size of $[\omega_L:\omega_R] \times [\omega_D:\omega_T]$. The inner finer grids $\Omega^2$ and $\Omega^3$ provide a better numerical resolution in the inner regions of the computational domain. The physical size of each nested grid in the $x$- and $y$-coordinate directions is determined as $(\omega_R-\omega_L)/2^{m-1}$ and $(\omega_T-\omega_B)/2^{m-1}$, where $m$ takes values from 1 to 3 in our simple example. Furthermore, each nested grid is subdivided into $N=4$ square cells. Such a mesh refinement is often needed to better resolve the central denser regions of astrophysical objects (e.g., gravitationally collapsing interstellar clouds of gas).  

The outside-in solution procedure consists of a sequence of steps starting from the calculation of the gravitational potential on the coarsest $\Omega^{m=1}$ grid. We discretize the Laplace operator using a simple five-point stencil
\begin{equation}
\label{poisson:descrete}
    \triangle \Phi \approx {1\over dx_m^2} \left( \Phi^m_{i+1,j}  - 2 \Phi^m_{i,j} + \Phi^m_{i-1,j} \right) + {1\over dy_m^2}  \left(\Phi^m_{i,j+1} - 2 \Phi^m_{i,j} + \Phi^m_{i,j-1}   \right),
\end{equation}
where $dx_m$ and $dy_m$ are the sizes of computational cells in $x$- and $y$-coordinate direction defined as $dx_m=(\omega_R - \omega_L)/(2^{m-1} N)$ and $dy_m=(\omega_T - \omega_D)/(2^{m-1} N)$, respectively. We note that we adopt $dx_m=dy_m$ throughout the paper. A similar algorithm applies to the third $z$-coordinate dimension in practical tests. 
The discretization of the Laplace operator introduces boundary values of the potential $\Phi_{\rm B}$ for each nested domain. The corresponding boundary ghost zones for the coarsest $\Omega^1$ grid are shown in Figure~\ref{fig:1} as additional layers of gray-colored cells. These boundary values can be found using the multipole expansion of the Laplace equation $\triangle \Phi=0$ in spherical harmonics (note that by our model setup the volume density outside the computational domain is equal zero). The pertaining formulae can be found in, for example, \cite{Jackson-electrodynamics}. The higher-order harmonics are added until a desired accuracy in $\Phi_{\rm B}^{m=1}$ is reached. This method usually converges fast for compact density configurations when most of the mass is concentrated in the center of the computational domain, but may become computationally costly or even diverge for arbitrary mass distributions with large amounts of mass present near outer boundaries.    

Once the boundary values $\Phi_{\rm B}^{m=1}$ of the gravitational potential are found, the potential on the coarsest grid $\Phi^{\rm m=1}$ can be computed using one of the methods of linear algebra. In this work, we employ the conjugate gradient method described in, for example, \citet{Press1992}. As with any iterative method, a truncation error has to be established {\it a priori}. The completion of the iterative process in the conjugate gradient method occurs when the norm of the relative residuals falls below $\epsilon=10^{-6}$. In practice, we substitute the calculated grid values of the potential into the numerical scheme for approximating the Poisson equation (see eq.~\ref{poisson:descrete}) and use the resulting data to form a residual difference vector with the grid values of the right-hand side ($4 \pi G \rho$). The norm of this residual vector should not exceed $10^{-6}$ relative to the norm of the right-hand side vector. It is important to note that the achieved accuracy is not the value of the relative error in the gravitational potential. The relative error in the gravitational potential is improved only when the grid is refined. In the iterative process, it is only guaranteed that the grid values of the potential, when substituting into the scheme for the Poisson equation, restore the right-hand side with a chosen truncation error of $\epsilon=10^{-6}$. The dependence of the solution on $\epsilon$ is investigated in Sect.~\ref{Compare}. The potentials $\Phi^{m}$ on the finer grids can then be found by taking the corresponding potentials on the coarser grids $\Phi^{\rm m-1}$ as the boundary values (which are known from the previous step). Moving in this manner from the outer coarsest grid toward the inner finest grid the gravitational potential in the entire computational domain can be computed. In the following text, the outlined solution procedure is denoted as the outside-in conjugate gradient method (hereafter, OiCG method).



\begin{figure}[h]
\centering
\includegraphics[scale=0.5]{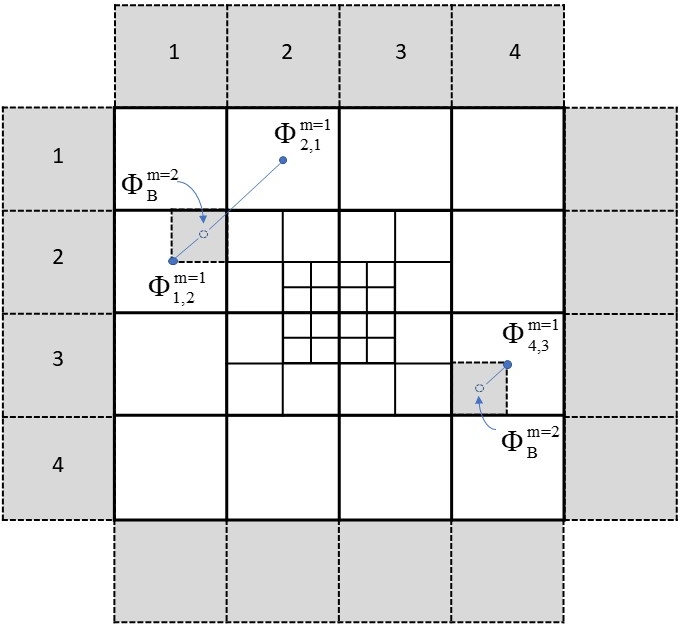}
\caption{Examples of finding the boundary potentials for the $m=2$ nested grid. The bottom-right example presents a simple extrapolation of the nearest potential at the $m=1$ grid. The top-left example shows a more sophisticated linear interpolation scheme, which involves two neighboring potentials at the $m=1$ level. 
}
\label{fig:2}
\end{figure}

An important prerequisite for the OiCG method is a proper interpolation of the potentials on a coarser grid to be used as the boundary values on a finer grid \citep{Matsumoto2003,Wang2020}. Figure~\ref{fig:2} shows two examples for the choice of the boundary values $\Phi^{\rm m=2}_{\rm B}$ on the second nested grid $m=2$. The first choice shown in the bottom-right corner consists in taking the nearest value of the potential on the corresponding coarser grid $m=1$, so that $\Phi^{\rm m=2}_{\rm B}=\Phi_{4,3}^{m=1}$. The subscripts (4,3) refer to the cell number on the two-dimensional $(x,y)$ mesh. This choice, however, is not optimal and will lead to degraded accuracy of the solution, as we subsequently demonstrate below. A better accuracy can be achieved by taking a linear interpolation of the potentials on the coarser $m=1$ grid to the center of the corresponding ghost zone of the $m=2$ mesh. The algorithm is illustrated in the upper-left corner of Figure~\ref{fig:2} and for the left-hand side interface between the nested grids ($i=1$, $x$-coordinate) can be described as follows
\begin{eqnarray}
\Phi^{\rm m+1}_{{\rm B},(i,k,l)}&=& 0.75 \, \Phi^{\rm m}_{ib,kb,lb}+0.25\, \Phi^{\rm m}_{ib+1,kb+1,lb+1} \, , \,\, \nonumber \label{eq:bound1} \\ &\,&\mathrm{if \, mod(k,2)}\ne 0 \,\, \mathrm{and \, mod(l,2)}\ne 0, \\
\Phi^{\rm m+1}_{{\rm B},(i,k,l)}&=& 0.75 \, \Phi^{\rm m}_{ib,kb,lb}+0.25\, \Phi^{\rm m}_{ib+1,kb+1,lb-1} \, , \,\, \nonumber \\ &\,&\mathrm{if \, mod(k,2)}\ne 0 \,\, \mathrm{and \, mod(l,2)}=0, \\
\Phi^{\rm m+1}_{{\rm B},(i,k,l)}&=& 0.75 \, \Phi^{\rm m}_{ib,kb,lb}+0.25\, \Phi^{\rm m}_{ib+1,kb-1,lb+1} \, , \,\, \nonumber \\ &\,&\mathrm{if \, mod(k,2)}=0 \,\, \mathrm{and \, mod(l,2)}\ne 0, \\
\Phi^{\rm m+1}_{{\rm B},(i,k,l)}&=& 0.75 \, \Phi^{\rm m}_{ib,kb,lb}+0.25\, \Phi^{\rm m}_{ib+1,kb-1,lb-1} \, , \,\, \nonumber \\ &\,&\mathrm{if \, mod(k,2)}=0 \,\, \mathrm{and \, mod(l,2)}=0. \label{eq:bound2}
\label{Eq:interpol}
\end{eqnarray}
Here, the indices $k$ and $l$ take values from 1 to $N+2$ and $ib$, $kb$, and $lb$ are related to $i$, $k$, and $l$ as $ib=N/4+i/2+1$, $kb=N/4+k/2+1$, and $lb=N/4+l/2+1$, respectively.
The algorithm can be generalized to other grid interfaces in a similar manner.
More sophisticated methods involving the introduction of additional boundary layers can be found in, e.g., \citet{Wang2020}.

\begin{figure*}
\centering
\includegraphics[scale=0.4]{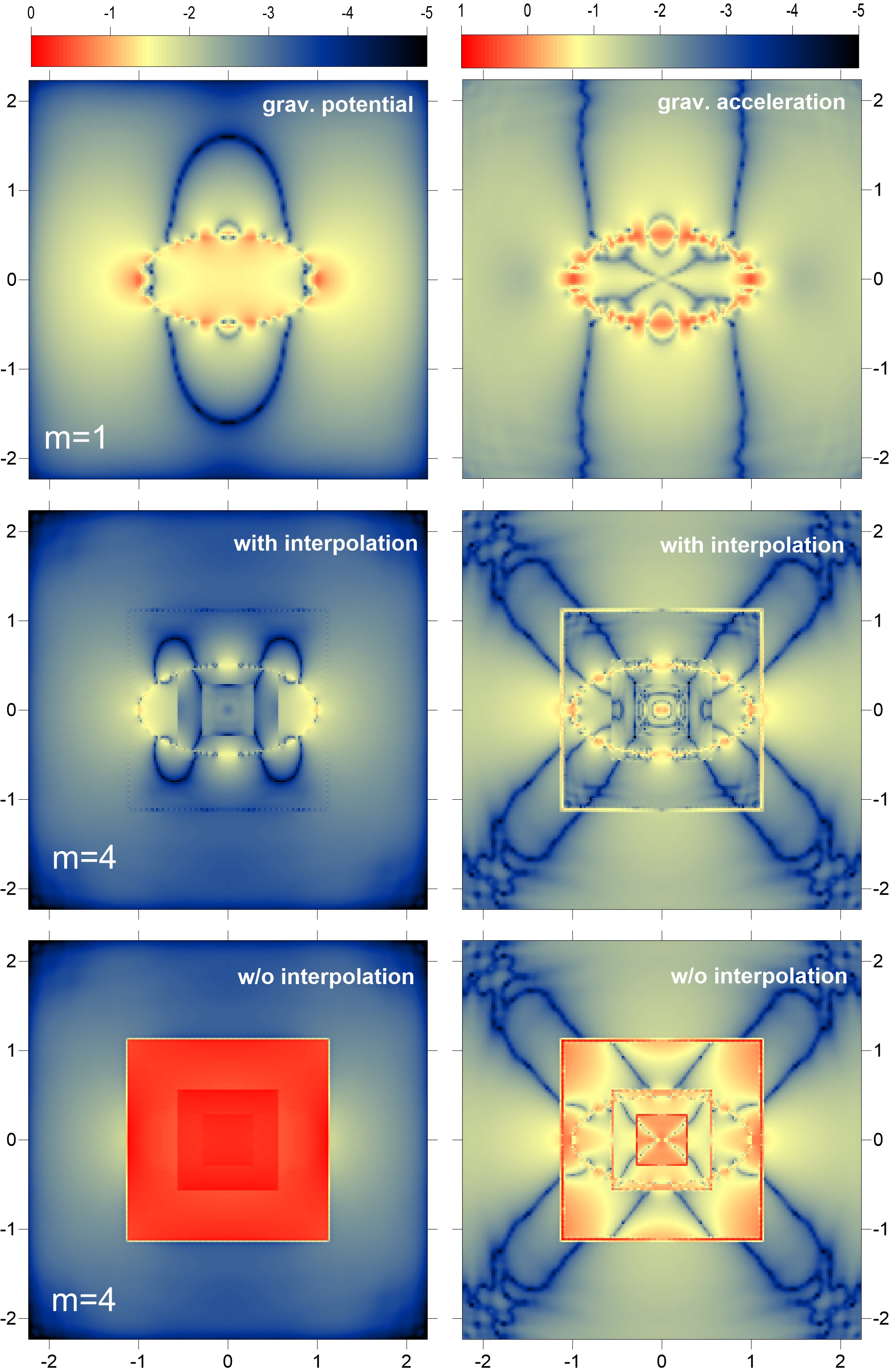}
\includegraphics[scale=0.4]{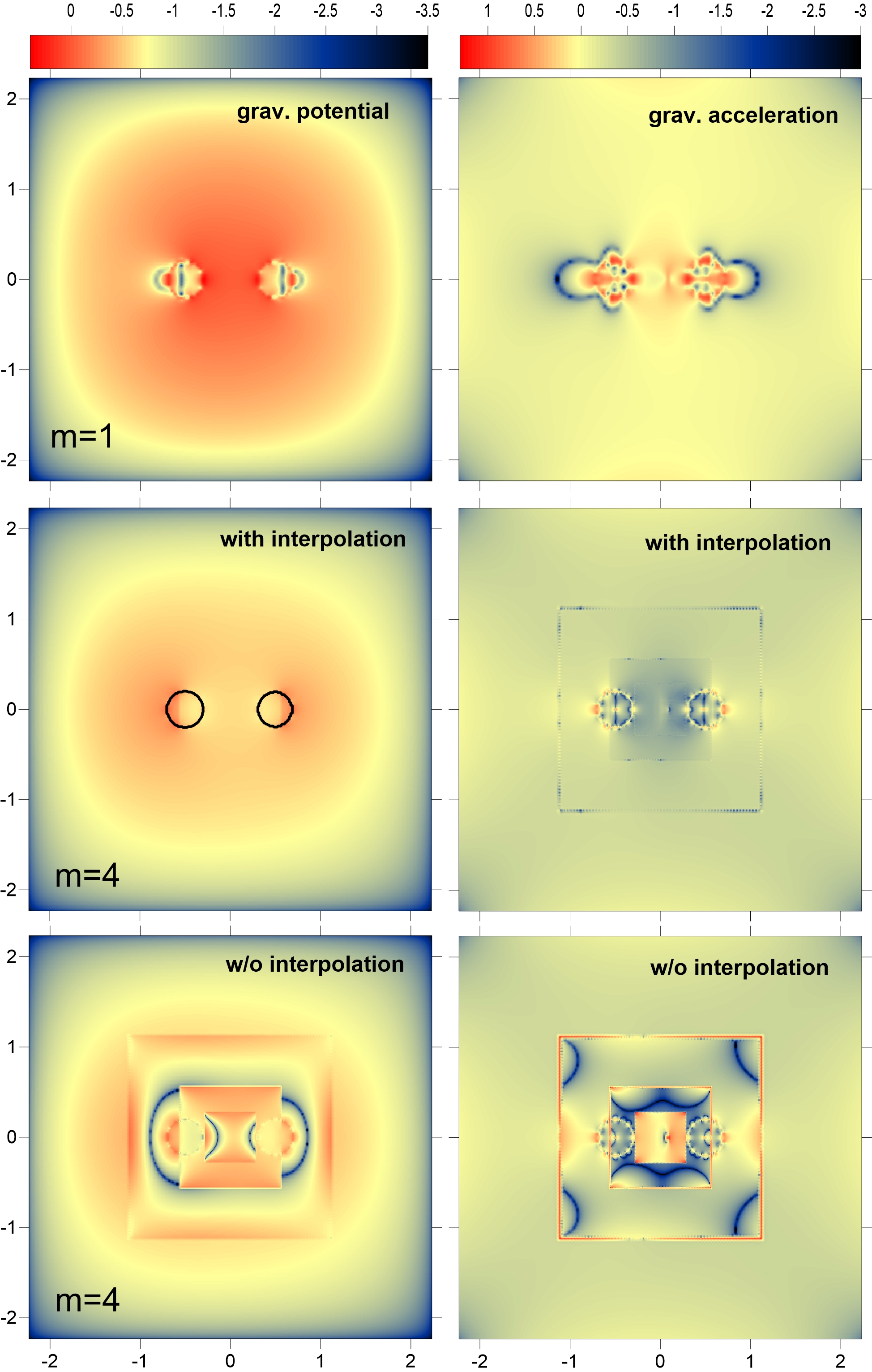}
\caption{Gravitational potential and acceleration on nested grids in the OiCG method. {\bf Left pair of columns:} Relative error in the calculation of the gravitational potential (first column) and gravitational acceleration (second column) for an oblate ellipsoid with the semi-major axis ratio 0.5:1.0:1.0. The horizontal bars present the relative errors in percent (log units). The top row illustrates the errors obtained with no mesh refinement ($m=1$). The middle and bottom rows show the errors for the case of four nested grids ($m=4$). In particular, the middle row corresponds to a linear interpolation of the boundary potentials, whereas the bottom row presents the errors when a simple extrapolation is used. {\bf Right pair of columns:} Relative error in the gravitational potential (first column) and gravitational acceleration (second column) for a binary star, the spatial location of which is outlined by two circles. The left component is twice as massive as the right one. 
}
\label{fig:3}
\end{figure*}

To test the accuracy of the OiCG method, we computed the gravitational potential and gravitational acceleration for two mass configurations with known analytical solutions: an oblate homogeneous ellipsoid and two nonintersecting homogeneous spheres. The former has a ratio of semi-major axes $a_1:a_2:a_3=0.5:1.0:1.0$. The latter is composed of two homogeneous spheres, both having a radius of 0.2. Their centers lie in the $x-y$ plane at -0.5 and 0.5 with respect to the coordinate center. This latter test model mimics a binary star. The computational domain has dimensions of [-2.25:2.25] in each coordinate direction. The ratio of masses of the two spheres is 2:1. The analytic solution for the gravitational potential of an oblate ellipsoid was derived by \citet{Chandra1969} and is provided in \citet{StoneNorman1992}. The analytic solution for the binary star is the sum of analytic solutions for the two homogeneous spheres, which can also be found in \citet{StoneNorman1992}.

The left pair of columns in Figure~\ref{fig:3} presents the relative errors in the gravitational potential and gravitational acceleration for an ellipsoid with the ratio of semi-major axes $a_1:a_2:a_3=0.5:1.0:1.0$. The errors in the gravitational potential and acceleration are calculated with respect to the analytical solutions as follows 
\begin{equation}
err_\Phi = \left|{\Phi_{\rm num} - \Phi_{\rm an} \over \Phi_{\rm an}} \right|; \,\,\, err_{\rm g} = \left|{|g|_{\rm num} - |g|_{\rm an} \over |g|_{\rm an}} \right|,
\end{equation}
where the subscripts ``num'' and ``an'' refer to the numerical and analytical values, respectively.

The modulus of gravitational acceleration $|\bl g|$ was calculated using a simple six-point stencil for the central discretization scheme, so that the $x$-component of the gravitational acceleration is written as
\begin{equation}
    g_x =-{\partial \Phi \over \partial x} \approx {\Phi^m_{i+1,j,k}-\Phi^m_{i-1,j,k} \over 2\,dx_m}.
    \label{eq:g_x}
\end{equation}
The other $y$- and $z$-components are expressed accordingly. This simple six-point stencil has to be 
modified at the cells adjacent to the boundary between fine and coarse grids. When calculating the gravitational acceleration at the cells of a finer grid that are adjacent to a coarser grid, we use the interpolated values of the gravitational potential as described in Equations~(\ref{eq:bound1})--(\ref{eq:bound2}). The opposite case of the gravitational acceleration on a coarser grid is more subtle. A straightforward method would be to use the potentials in the adjacent cells on a finer grid, arithmetically averaged to the geometrical center of the underlying coarser cell. A more physically motivated method makes use of the Gauss theorem and requires that the gravitational acceleration at the coarse-fine grid interface be equal when computed using the fine and coarse grid cells, respectively. We review these cases in Appendix~\ref{sect:App1} and note that we used the first simple method in this paper but in real simulations the second method may be more appropriate. 
The boundary conditions on the gravitational potential for the outermost mesh $m=1$ were calculated using the analytical solution. Two mesh configurations were considered: $m=1$ (no mesh refinement) and $m=4$ (four nested grids). The number of grid cells for each mesh was set equal to $N=128$ in each coordinate direction. Clearly, the solution with a simple extrapolation of the boundary potentials to the nearest cell on the coarse grid (as shown in the bottom row of Fig.~\ref{fig:2}) produces a notably higher relative error in both $\Phi$ and $|\bl g|$ than the solution that employs a more sophisticated linear interpolation scheme (middle row in Fig.~\ref{fig:2}). In the latter case, the mesh interfaces can still be noticed as locations where the relative errors experience jumps or
discontinuities.  

The relative errors in the gravitational potential and gravitational acceleration for a binary star are shown in the right pair of columns in Figure~\ref{fig:3}. This example is useful because it tests the method on a nonaxisymmetric mass distribution. The resulting trends are nevertheless similar -- mesh refinement and linear interpolation of the boundary potentials help to improve the accuracy and reduce the errors.
The corresponding errors relative to the analytic solution for both density configurations, as well as the standard deviation in the relative errors $\sigma$, are provided in Tables~\ref{table:1} and \ref{table:2} for comparison. The numbers separated by the slash correspond to the cases without and with a linear interpolation following Eq.~(\ref{Eq:interpol}), respectively, and $\sigma$ represents the standard deviation in the errors.

\begin{table*}
\center
\caption{\label{table:1} Errors in gravitational potential: ellipsoid (left) and binary star (right)}
\begin{tabular}{cccc|ccc}
\hline 
\hline 
$m$ & max. err. & mean err. &  $\sigma$ & max. err. & mean err. &  $\sigma$ \tabularnewline
 & [\%] &  [\%] & [\%] & [\%] &  [\%] & [\%]   \tabularnewline
\hline 
1 &  0.25 & 0.01 &  0.017 & 1.6 & 0.08 & 0.16 \tabularnewline
4 &  0.85/0.06 & 0.42/0.003 &  0.25/0.0045 & 0.84/1.0 & 0.23/0.25 & 0.14/0.09 \tabularnewline
\hline 
\end{tabular}
\end{table*}

\begin{table*}
\center
\caption{\label{table:2} Errors in gravitational acceleration: ellipsoid (left) and binary (right)}
\begin{tabular}{cccc|ccc}
\hline 
\hline 
$m$ & max. err. & mean err. &  $\sigma$  & max. err. & mean err. &  $\sigma$  \tabularnewline
 & [\%] &  [\%] &  [\%] & [\%] &  [\%] &  [\%] \tabularnewline
\hline 
1 &  2.6 & 0.058 &  0.145 & 9.95 & 0.86 & 0.47  \tabularnewline
4 &  25/1.23 & 0.52/0.042 &  1.45/0.08 & 55.5/3.66 & 1.1/0.37 & 1.7/0.23  \tabularnewline
\hline 
\end{tabular}
\end{table*}

\section{Integral form of the gravitational potential and the convolution method}
\label{convolution}
An alternative approach to computing the gravitational potential is to use the integral representation 
\begin{equation}
 \Phi({\bf r}) = -G \iiint \limits_V  {\rho({\bf r}^\prime) \over |{\bf r} - {\bf r}^\prime|} d^3{\bf r}^\prime  ,
 \label{int:pot}
\end{equation}
where the integration is performed over the volume $V$ occupied by the computational domain. On a discrete Cartesian computational mesh the gravitational potential $\Phi(x_i,y_j,z_k)$ in a cell $(i,j,k)$ can be written in the following form
\begin{eqnarray}
    \Phi(x_i,y_j,z_k) &=& -G \sum_{i^\prime,j^\prime,k^\prime} { M(x_{i^\prime},y_{j^\prime},z_{k^\prime}) \over \sqrt{(x_i-x_{i^\prime})^2 + (y_j-y_{j^\prime})^2 + (z_k-z_{k^\prime})^2} } \nonumber \\
   &=& - G \sum_{i^\prime,j^\prime,k^\prime} {M(x_{i^\prime},y_{j^\prime},z_{k^\prime}) \, {\cal G}(x_i-x_{i^\prime},y_j-y_{j^\prime},z_k-z_{k^\prime} ) } ,
    \label{sum:pot}
\end{eqnarray}
where $M(x_{i^\prime},y_{j^\prime},z_{k^\prime})$ is the mass in a numerical cell ($i^\prime,j^\prime,k^\prime$) with coordinates ($x_{i^\prime},y_{j^\prime},z_{k^\prime}$), ${\cal G}(x_i-x_{i^\prime},y_j-y_{j^\prime},z_k-z_{k^\prime})$ is the inverse distance between cells $(i,j,k)$ and ($i^\prime,j^\prime,k^\prime$), and the summation is performed over all numerical cells. 

Finding the gravitational potential for a numerical mesh with $N^3$ cells by direct summation requires $(2N)^6$ operations, which becomes numerically expensive for $N$ greater than a few tens. Therefore, the convolution theorem is often invoked to compute the potential given by Eq.~(\ref{sum:pot}). The convolution theorem states that the sum of the following form \citep{BT1987}
\begin{equation}
h_k= {1 \over \sqrt{2 K} }   \sum_{k^\prime=-N}^{k^\prime=N-1} f_{k-k^\prime} \, g_{k^\prime}
\label{sum:convolve}
\end{equation}
can be computed by taking the Fourier transforms separately for functions $f_{k-k^\prime}$ and $g_{k^\prime}$, multiplying them in the Fourier space to form $\hat{h}_p=\hat{f}_p \hat{g}_p$, and finally taking the inverse Fourier transform of $\hat{h}_p$. For a two-dimensional grid, for example, this requires $2N[6\log_2(2N)+1]$ additions and multiplications to be compared to $(2N)^2$ operations for a direct evaluation of the sum~(\ref{sum:convolve}), see \citet{BT1987}.
 We note that the classic Fourier transformation method to solve the Poisson equation \citep[e.g.,][]{Press1992} is inapplicable to Equation~(\ref{sum:pot}). This is because this method is used to turn a system of discretized Poisson equations into an algebraic expression in the Fourier space, whereas Equation~(\ref{sum:pot}) is already an algebraic expression, the value of which can be found most efficiently using the convolution theorem.

\begin{figure}[h]
\centering
\includegraphics[width=1\columnwidth]{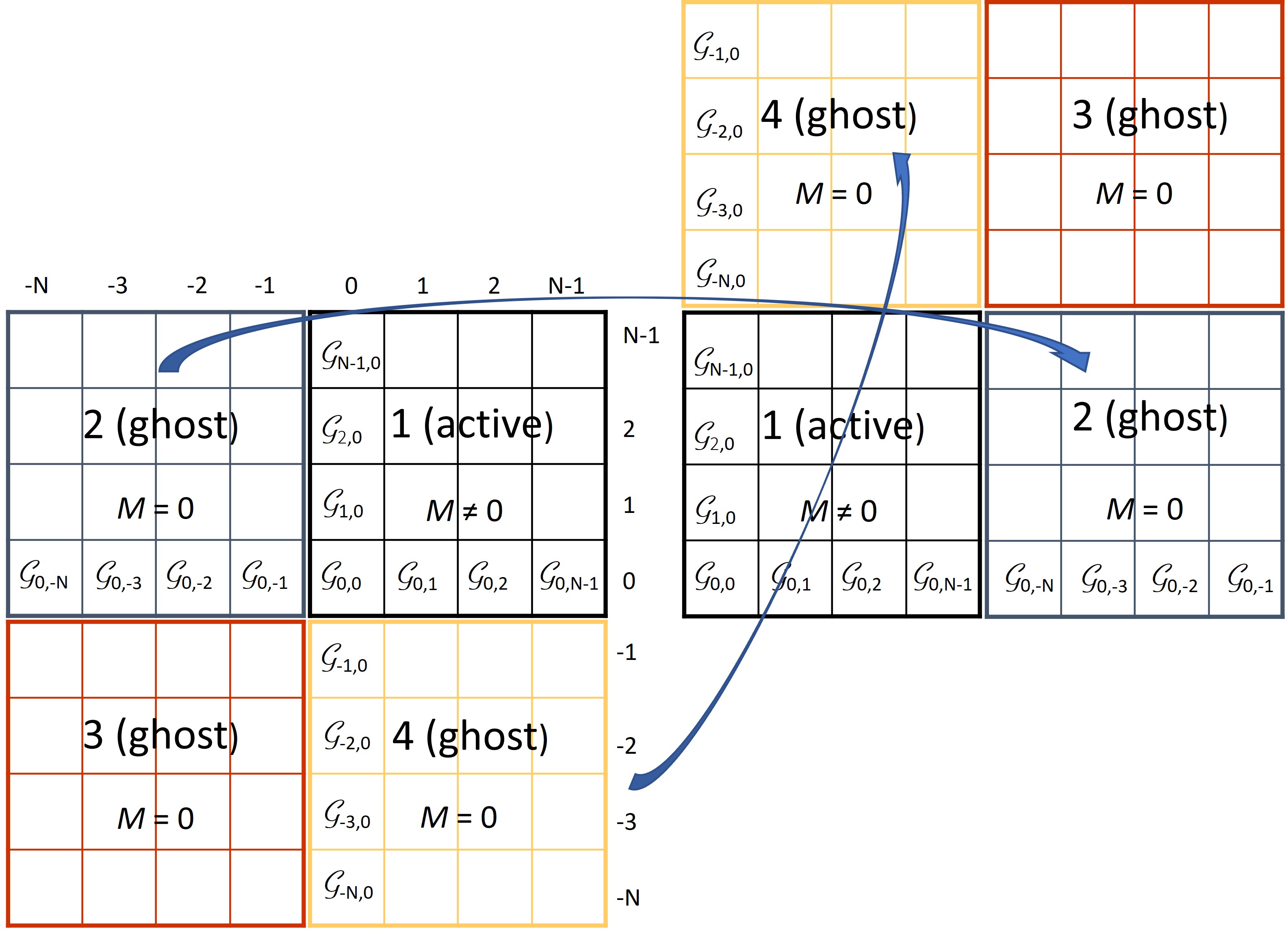}
\caption{Schematic illustration of the mesh doubling and rearrangement in the convolution method. The original active region 1 is doubled in each coordinate direction to form three ghost regions (2, 3, and 4) that are filled with zero masses ($M=0$). The doubled mesh is rearranged as shown by the arrows to make the inverse distance $G$ and mass $M$ periodic functions on the new domain.
}
\label{fig:4}
\end{figure}

To apply the convolution theorem to the gravitational potential given by Eq.~(\ref{sum:pot}), certain manipulations with the mass $M$ and inverse distance ${\cal G}$ have to be made. The convolution theorem is applicable to periodic functions, whereas $M$ and $\cal G$ may not be periodic. This problem can be solved by doubling the computational domain in each coordinate direction and assigning zero values to $M$ on this extended mesh (note that $M$ on the original domain is kept intact). Both $\cal G$ and $M$ can then be made periodic on the extended domain by rearranging the extended mesh as shown in Fig.~\ref{fig:4} \citep[see also, e.g.,][]{HE1988}. We note that the zero-padding of the ghost domains makes the mass distribution in the active domain effectively isolated, meaning that this method can be only applied to isolated mass distributions and not to periodic ones.   

Furthermore, one can notice that the inverse distance $\cal G$ formally turns into infinity when $i=i^\prime$, $j=j^\prime$, and $k=k^\prime$. This problem can be solved by introducing a smoothing length $\epsilon_{\rm sm}$, so that the inverse distance takes the following form 
\begin{equation}
    {\cal G}=  {1 \over \sqrt{(x_i-x_{i^\prime})^2 + (y_j-y_{j^\prime})^2 + (z_k-z_{k^\prime})^2 + \epsilon_{\rm sm}^2} }.
\end{equation}
However, there is always an uncertainty as to what value is to assign to $\epsilon_{\rm sm}$ \citep{MullerKley2012}. A way to circumvent the problem is to evaluate the input to the potential in the cell ($i=i^\prime$, $j=j^\prime$, $k=k^\prime$) from the mass distribution within the cell itself. If we assume that the density $\rho$ is constant within the cell, then the corresponding potential can be found as (see Eq.~\ref{int:pot})
\begin{eqnarray}
 \Phi(x_i&=&x_{i^\prime},y_j=y_{j^\prime},z_k=z_{k^\prime})=  \nonumber \\
 &=& - G \, \rho(x_i,y_j,z_k) \int \limits_{-{dx \over 2}}^{dx \over 2}  \int \limits_{-{dy \over 2}}^{dy \over 2} \int \limits_{-{dz \over 2}}^{dz \over 2}  {dx^\prime dy^\prime dz^\prime \over \sqrt{x^{\prime2} + y^{\prime2} +z^{\prime2}} },
 \label{pot:cell}
\end{eqnarray}
where $dx$, $dy$, and $dz$ are the sizes of the cell in each coordinate direction. We also used variable substitutions of the following form: $x_i-x_{i^\prime}=x^\prime$, $y_j-y_{j^\prime}=y^\prime$, and $z_k-z_{k^\prime}=z^\prime$. The integral in Eq.~(\ref{pot:cell}) can be found either numerically or analytically \citep{Macmillan1958,Hahn2020}. We note that the assumption of constant $\rho$ may be relaxed and the input to the potential from the cell itself can also be computed for piecewise-linear density profiles \citep{Hahn2020}.

We emphasize that the convolution method does not require us to precompute the boundary values of the gravitational potential at the coarsest grid (using, e.g., the multipole expansion) because of the absence of spatial derivatives in Eq.~(\ref{sum:pot}). This can be considered as a clear advantage compared to numerical solutions of the Poisson equation. Nevertheless, the practical applicability of the method described in this section has been limited to Cartesian meshes with grid cells of similar size in each coordinate direction (equidistant grids). This limitation is dictated by the fast Fourier transform (FFT), which was developed for the case of equidistant grids. There are extensions of the FFT methods to nonequidistant grids \citep[e.g.,][]{Potts2004}, but in our experience they usually work notably slower than the classic FFT methods. The convolution method can be also applied to finding the gravitational potential of disk-like configurations on the polar grid ($r,\phi$) with a logarithmic scaling in the radial direction. In this case, the introduction of a new variable $u=\ln r$ helps to map the polar grid onto the equidistant Cartesian grid, so that the classic convolution theorem can be used \citep{BT1987}. However, such cases are rather specific and for a three-dimensional cylindrical grid ($z,r,\phi$) the above described method of variable substitution does not work. For 3D cylindrical coordinates, the convolution theorem can be applied only along the $\phi$ and $z$-directions, while in the radial direction the direct summation has to be used, which considerably decelerates the method. For 3D spherical coordinates the situation is even worse since the convolution theorem works only along the $\phi$-coordinate.

\begin{figure}[h]
\centering
\includegraphics[width=1\columnwidth]{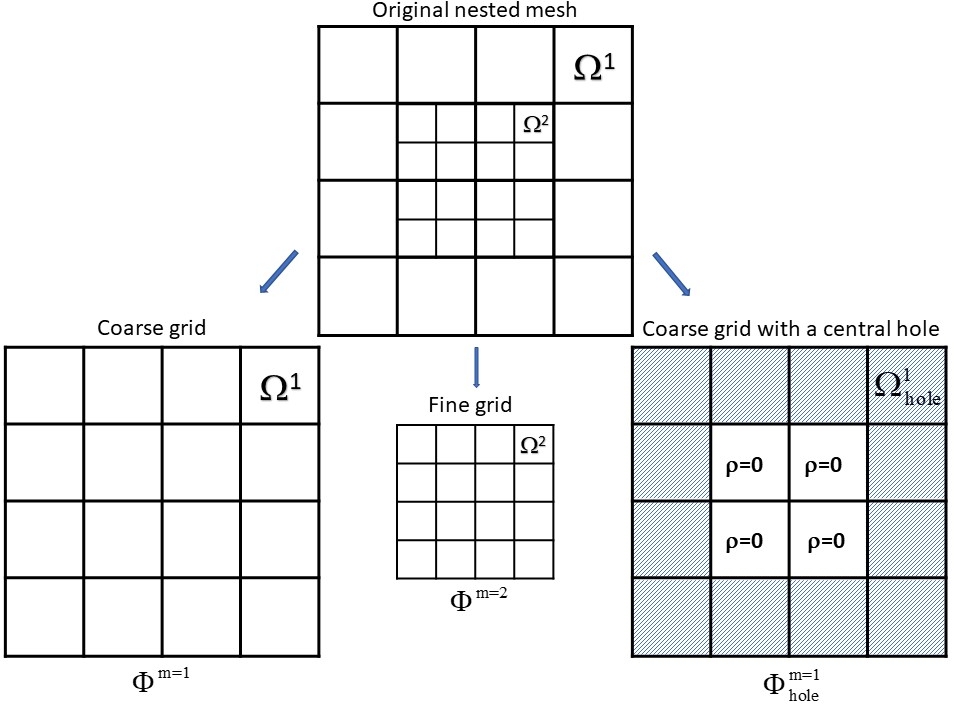}
\caption{Decomposition of two nested grids (top) into three individual grids with cells of equal size to which the classic FFT method can be applied: a coarse grid ($\Omega^1$, bottom left), refined grid ($\Omega^2$, bottom center), and coarse grid with a cavity ($\Omega^1_{\rm hole}$, bottom right).}
\label{fig:5}
\end{figure}

\section{Modification of the convolution method for nested grids}
\label{nested}
In this section, we describe a modification to the convolution method that can be applied to nested Cartesian grids. A nested grid with only two levels of refinement ($\Omega^1:\Omega^2$) is used for simplicity to illustrate the method but the procedure can easily be generalized to any level of refinement.

Figure~\ref{fig:5} presents an example of the mesh decomposition, which is used to compute the gravitational potential on nested grids. The original mesh (top) is split into three regular grids to which the FFT method can be applied: a coarse $m=1$ grid ($\Omega^1$, bottom left), a refined $m=2$ grid ($\Omega^2$, bottom center), and a coarse $m=1$ grid with a central cavity ($\Omega^1_{\rm hole}$, bottom right). The physical size of the cavity is equal to that of the refined $m=2$ grid. We note that the density $\rho$ in the cavity is set equal to zero. The grid with a central cavity accounts for the gravitational potential in the refined $m=2$ grid created by the matter that lies in the hatched cells outside of this grid. Indeed, if we calculate the potentials on the coarse and refined grids, and then simply add them up, the procedure would not be complete. This is because the calculation of the potential on the refined grid does not take into account the contribution of the matter that lies outside this grid.

\begin{figure}[h]
\centering
\includegraphics[scale=0.5]{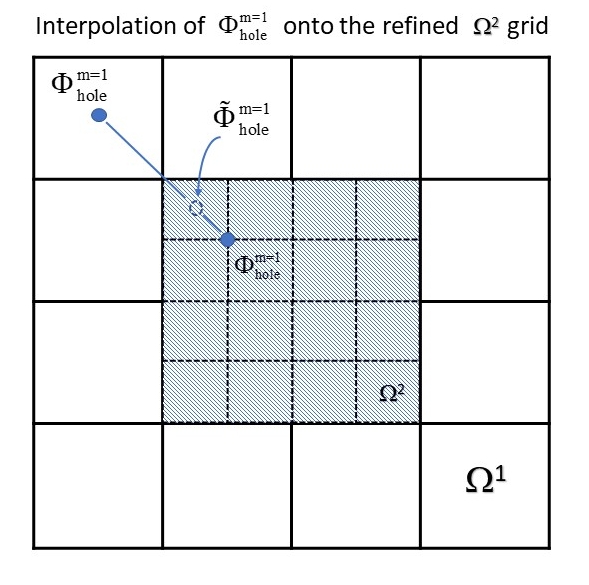}
\caption{Interpolation of the gravitational potential $\Phi^{\rm m=1}_{\rm hole}$, obtained on the coarse $m=1$ grid with a central cavity, on the refined $m=2$ grid highlighted with hatches. The resulting potential $\tilde{\Phi}^{\rm m=1}_{\rm hole}$ is shown with the arrow.
}
\label{fig:6}
\end{figure}

The solution procedure consists of three steps: 1) calculating the potential $\Phi^{\rm m=2}$ on the refined $m=2$ grid, 2) calculating the potential $\Phi^{\rm m=1}$ on the coarse $m=1$ grid, and 3) calculating the potential $\Phi^{\rm m=1}_{\rm hole}$ on the coarse $m=1$ grid with a central cavity. An additional substep includes an interpolation of $\Phi^{\rm m=1}_{\rm hole}$ to the centers of the refined $m=2$ grid as, for example, shown in Figure~\ref{fig:6}. The resulting interpolated potential $\tilde{\Phi}^{\rm m=1}_{\rm hole}$ is used in the construction of the final solution. Other more sophisticated interpolation procedures can also be devised and may improve the accuracy of the final solution.

\begin{figure}[h]
\centering
\includegraphics[width=1\columnwidth]{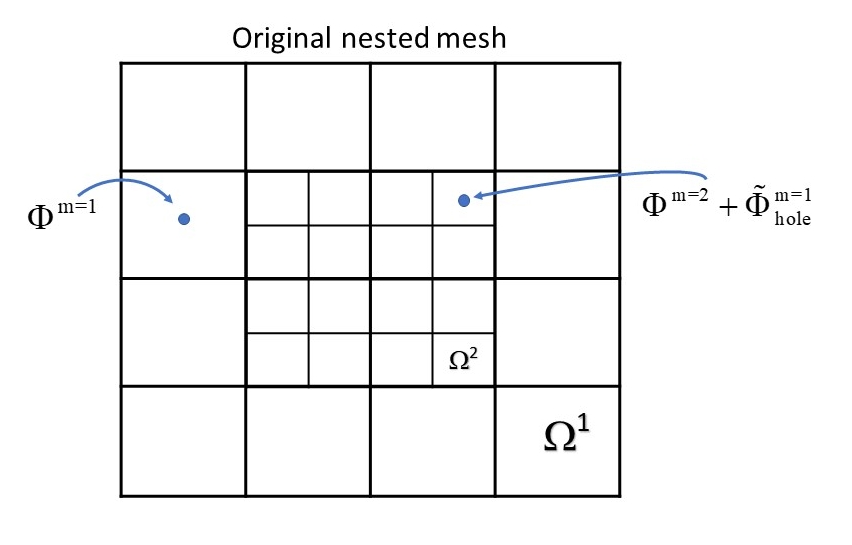}
\caption{Construction of the final solution on the original nested mesh ($\Omega^1:\Omega^2$).
}
\label{fig:7}
\end{figure}

The potential on the original nested mesh is then constructed as illustrated in Figure~\ref{fig:7}. The gravitational potential on the coarse $m=1$ grid is set equal to $\Phi^{\rm m=1}$, while the potential on the refined $m=2$ grid is set equal to the sum of $\Phi^{m=2}$ and $\tilde{\Phi}^{\rm m=1}_{\rm hole}$. 
We note that each of the three steps is independent and can be done in parallel, unlike the outside-in solution procedure outlined in Sect.~\ref{inside-out}. This procedure can be extended to any number of nested grids as follows
\begin{eqnarray}
\label{eq:meshes}
    \Phi(\Omega^1) &=& \Phi^{\rm m=1}, \nonumber \\
    \Phi(\Omega^2) &=& \Phi^{\rm m=2} +  \tilde{\Phi}^{\rm m=1}_{\rm hole}, \nonumber \\
    \Phi(\Omega^3) &=& \Phi^{\rm m=3} + \tilde{\Phi}^{\rm m=1}_{\rm hole} + \tilde{\Phi}^{\rm m=2}_{\rm hole}, \nonumber \\
    \Phi(\Omega^n) &=& \Phi^{m=n} + \sum_{k=1}^{n-1}  \tilde{\Phi}^{m=k}_{\rm hole},
\end{eqnarray}
where $\Phi(\Omega^n)$ is the gravitational potential on the $n$-th nested grid. In the following section, we use four nested grids to test the proposed convolution method.

\section{Comparison of the models}
\label{Compare}
In this section, we carry out a comparison of the gravitational potential and acceleration calculated using the two considered models: the convolution method and OiCG method. 
As in Sect.~\ref{inside-out}, we consider an oblate ellipsoid and a binary star. We note, however, that here we use the multipole expansion with up to 6 multipole moments to find the gravitational potential at the outer boundary of the coarsest $m=1$ grid. This makes the comparison more realistic because in real simulations the analytic solution is rarely available. In both cases, linear interpolation of the boundary potentials at the interfaces between the nested grids is applied. In the convolution method, the interpolation is only used to obtain the gravitational acceleration, while in the OiCG method the interpolation is also required to calculate the boundary potentials at the finer grids for the outside-in procedure to work.

\begin{figure*}
\centering
\includegraphics[scale=0.4]{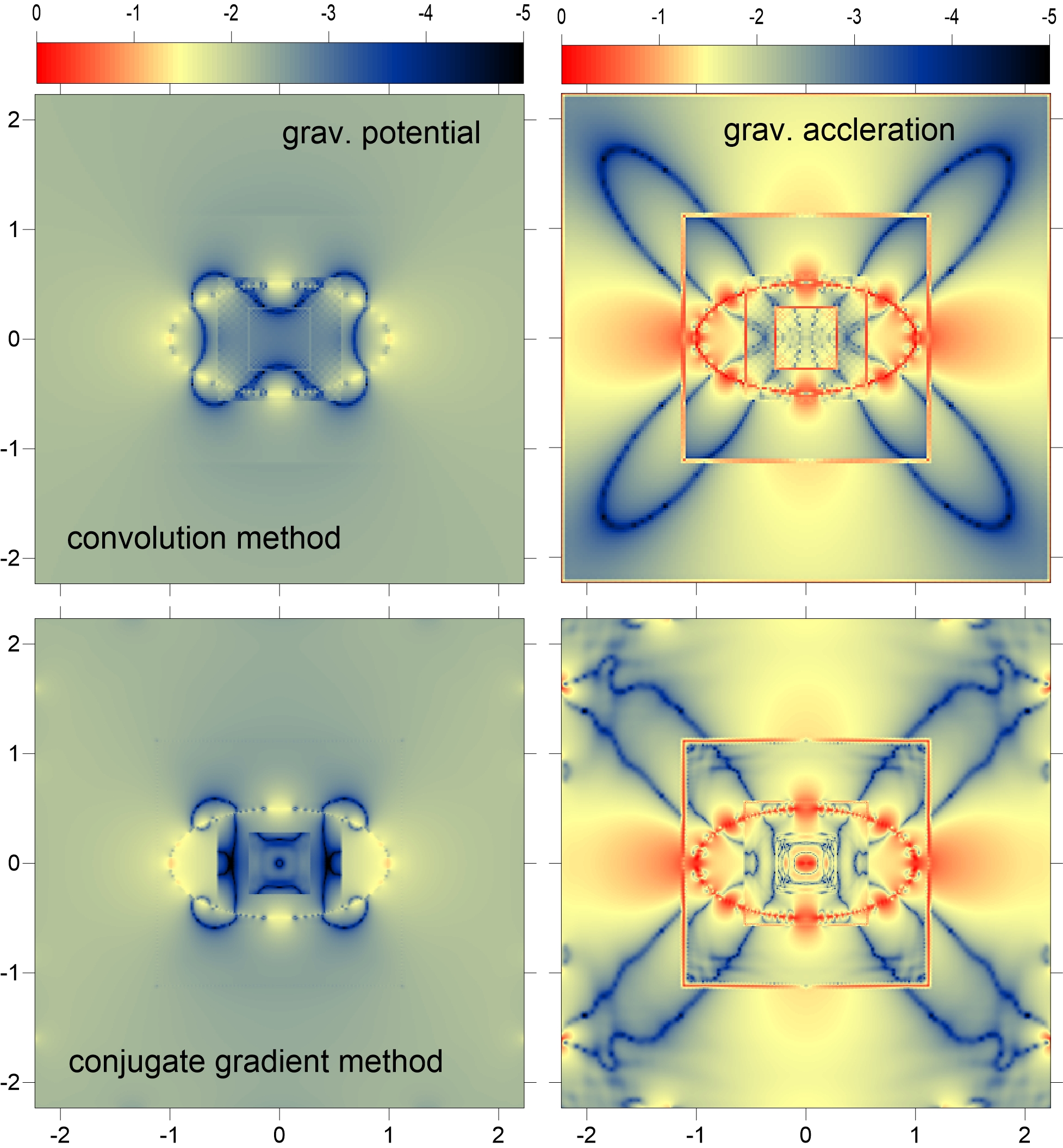}
\includegraphics[scale=0.4]{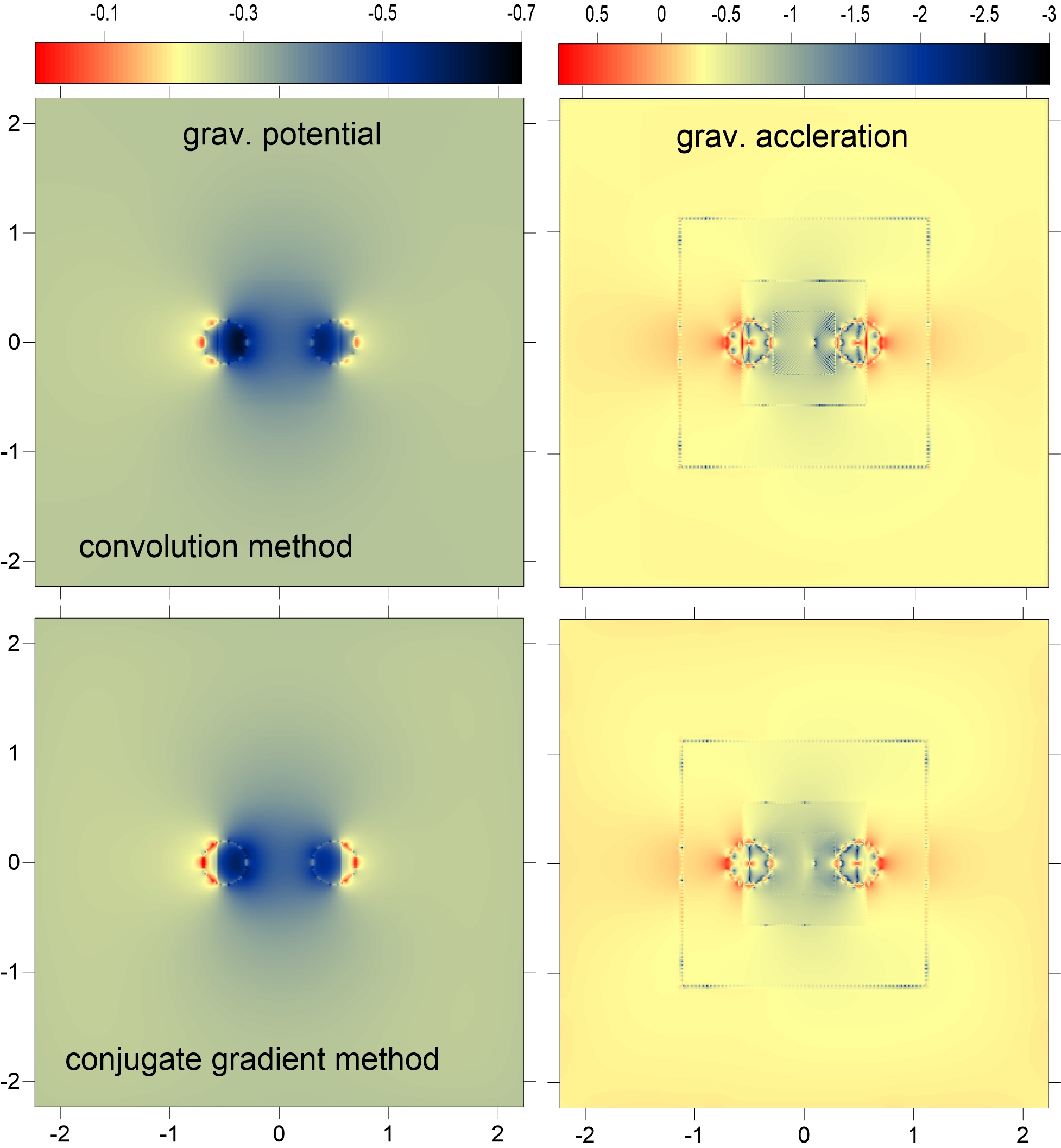}
\caption{ Comparison of the relative errors in the gravitational potential and acceleration for an oblate ellipsoid (left pair of columns) and binary star (right pair of columns).
The top row shows the results for the convolution method, while the bottom row presents the relative errors for the OiCG method. Four nested grids and 128 grid zones in each coordinate direction are used.}
\label{fig:10}
\end{figure*}

We found that both methods are characterized by comparable accuracy, as can be seen from Figure~\ref{fig:10} for the case of four nested grids. The relative errors are provided in Tables~\ref{table:3} and \ref{table:4} for convenience. For instance, the maximum error in the gravitational potential does not exceed 1.0\% in the case of the binary star and is notably smaller for the ellipsoid. The mean errors are a few per mille (in \%) in the case of the ellipsoid and just a fraction of a percent for the binary. As expected, the calculation of the gravitational acceleration shows larger relative errors and the maximum values can reach 10\% for the binary. Nevertheless, the mean errors are smaller than 1\%. The standard deviation is also low, implying that large errors are rare.  
A closer look at Figure~\ref{fig:10} indicates that the interfaces between the nested grids are hardly seen in the errors of the gravitational potential, whereas the errors in the gravitational acceleration are often the highest at the position of the interfaces. This means that more accurate schemes for calculating the gravitational acceleration from the gravitational potential (rather than a simple six-point stencil) may be needed to reduce the error and improve the accuracy \citep[e.g.,][]{Wang2020}.    

\begin{table*}
\center
\caption{\label{table:3} Errors in gravitational potential: ellipsoid (left) and binary star (right)}
\begin{tabular}{cccc|ccc}
\hline 
\hline 
method & max. err. & mean err. &  $\sigma$ & max. err. & mean err. &  $\sigma$ \tabularnewline
 & [\%] &  [\%] & [\%] & [\%] &  [\%] & [\%]   \tabularnewline
\hline 
conj. gr. &  0.06 & 0.003 &  0.0045 & 1.0 & 0.25 & 0.09 \tabularnewline
convolution &  0.045 & 0.004 &  0.004 & 0.85 & 0.43 & 0.078 \tabularnewline
\hline 
\end{tabular}
\end{table*}

\begin{table*}
\center
\caption{\label{table:4} Errors in gravitational acceleration: ellipsoid (left) and binary (right)}
\begin{tabular}{cccc|ccc}
\hline 
\hline 
method & max. err. & mean err. &  $\sigma$  & max. err. & mean err. &  $\sigma$  \tabularnewline
 & [\%] &  [\%] &  [\%] & [\%] &  [\%] &  [\%] \tabularnewline
\hline 
conj. gr. &  1.23 & 0.04 &  0.08 & 3.66 & 0.37 & 0.23  \tabularnewline
convolution &  1.74 & 0.05 &  0.12 & 9.4 & 0.38 & 0.29  \tabularnewline
\hline 
\end{tabular}
\end{table*}

\begin{figure}
\centering
\includegraphics[width=1\columnwidth]{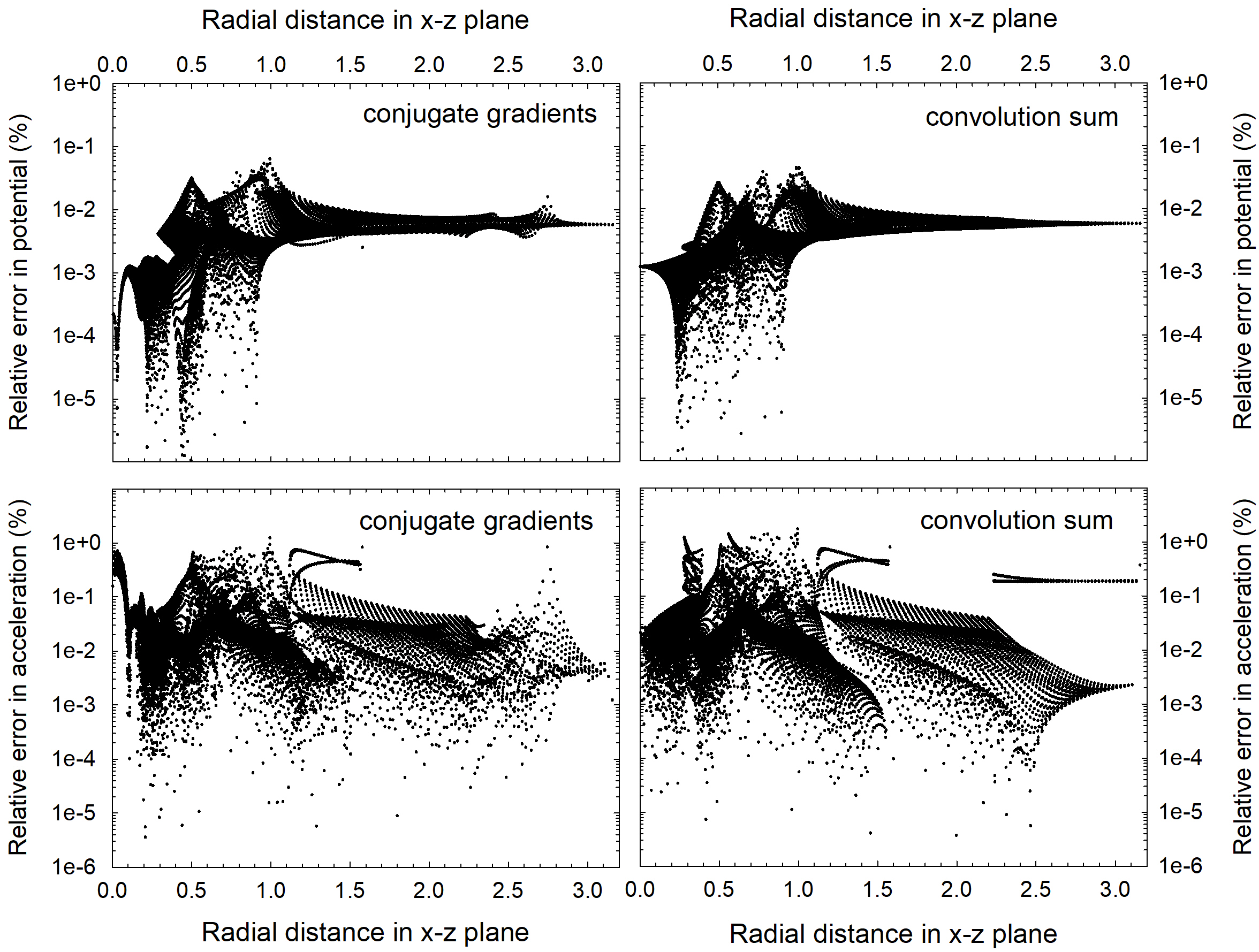}
\caption{Relative errors in the gravitational potential (top row) and gravitational acceleration (bottom row) as a function of radial distance from the coordinate center in the vertical $x-z$ plane. The case of the oblate ellipsoid is shown. Left and right columns present the results for the OiCG and convolution method, respectively.
}
\label{fig:11}
\end{figure}

Figure~\ref{fig:11} presents the relative errors (in \%) for the oblate ellipsoid as a function of radial distance from the coordinate center in the vertical $x-z$ plane. The errors have a wide spread but are mostly below 0.01\% for the potential and 0.1\% for the acceleration. There is little systematic difference between the two considered methods.

\begin{figure}
\centering
\includegraphics[width=1\columnwidth]{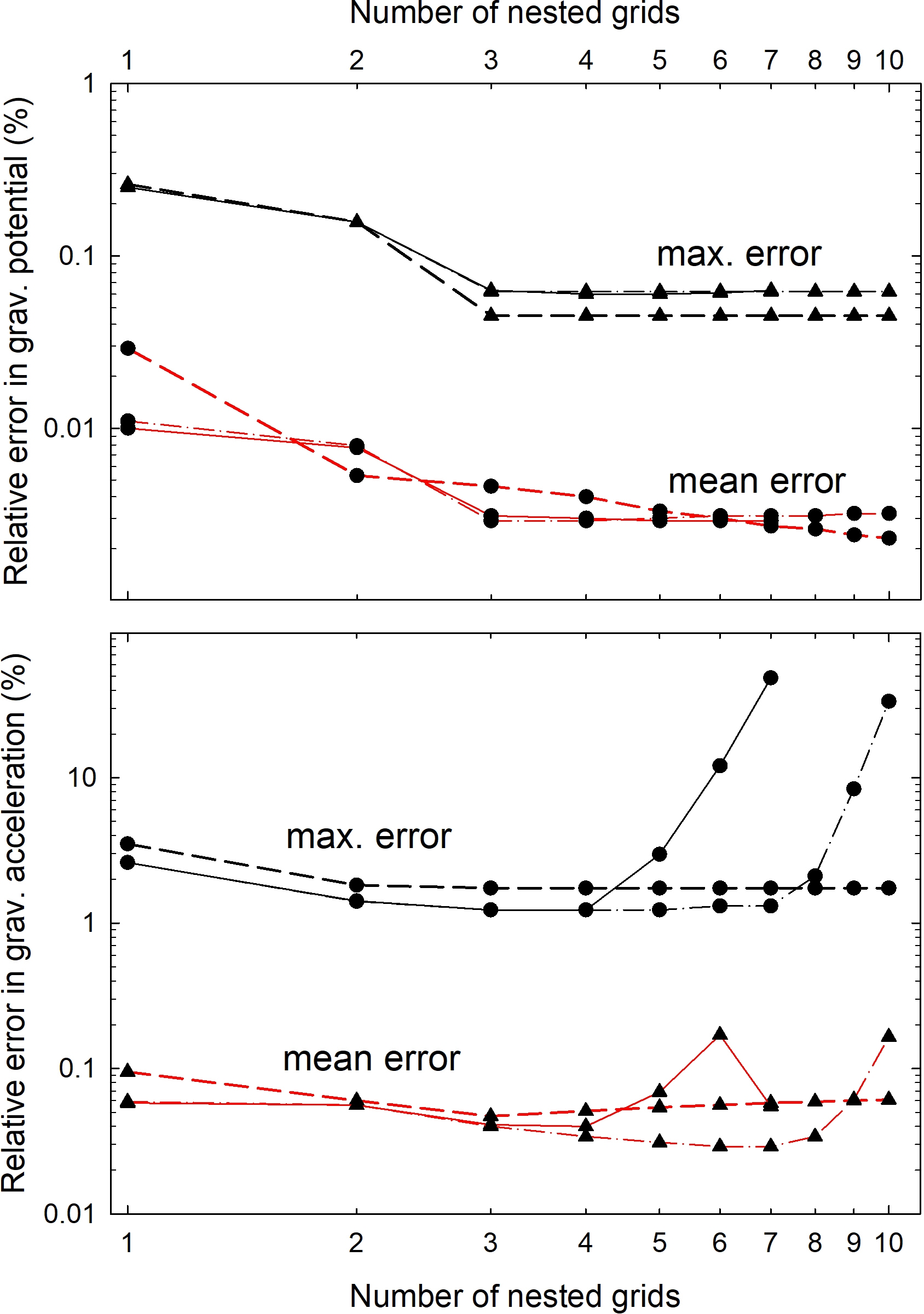}
\caption{Relative errors for the gravitational potential (top panel) and gravitational acceleration (bottom panel) for the oblate ellipsoid as a function of the number of nested grids. The black and red lines present the maximum and mean errors, respectively. The dashed lines correspond to the convolution method, while the solid and dash-dotted lines show the results for the OiCG method with the tolerance $\epsilon$ set equal to $10^{-6}$ and $10^{-8}$, respectively.
}
\label{fig:12}
\end{figure}

As another comparison test, we calculate the maximum and mean errors in the gravitational potential and acceleration of the oblate ellipsoid for nested grids ranging from $m=1$ to $m=10$. 
The results are shown in Figure~\ref{fig:12}. The errors in the vertical $x-z$ plane passing through the coordinate center are only considered. In the case of the OiCG method, we also vary the tolerance $\epsilon$ with which the solution is found by setting it to $10^{-6}$ and $10^{-8}$ (see Sect.~\ref{inside-out}). We note that the convolution method is not iterative and this procedure is not needed. Figure~\ref{fig:12} indicates that both methods are characterized by errors in the gravitational potential of comparable magnitude. The convolution method produces relative errors in $\Phi$ that decline with increasing $m$, while the corresponding errors in the OiCG method saturate already at $m\ge3$. An important difference between the two considered methods can be seen in the behavior of relative errors for the gravitational acceleration. While the errors in the convolution method saturate with increasing $m$, the errors in the OiCG method begin to grow. The turning point depends on the tolerance $\epsilon$ with which the iterations are controlled. We found that the deterioration occurs in the very central regions of the ellipsoid. This test problem reveals a potentially dangerous behavior of the OiCG method and requires further investigation in the future. 

\begin{figure}
\centering
\includegraphics[width=1\columnwidth]{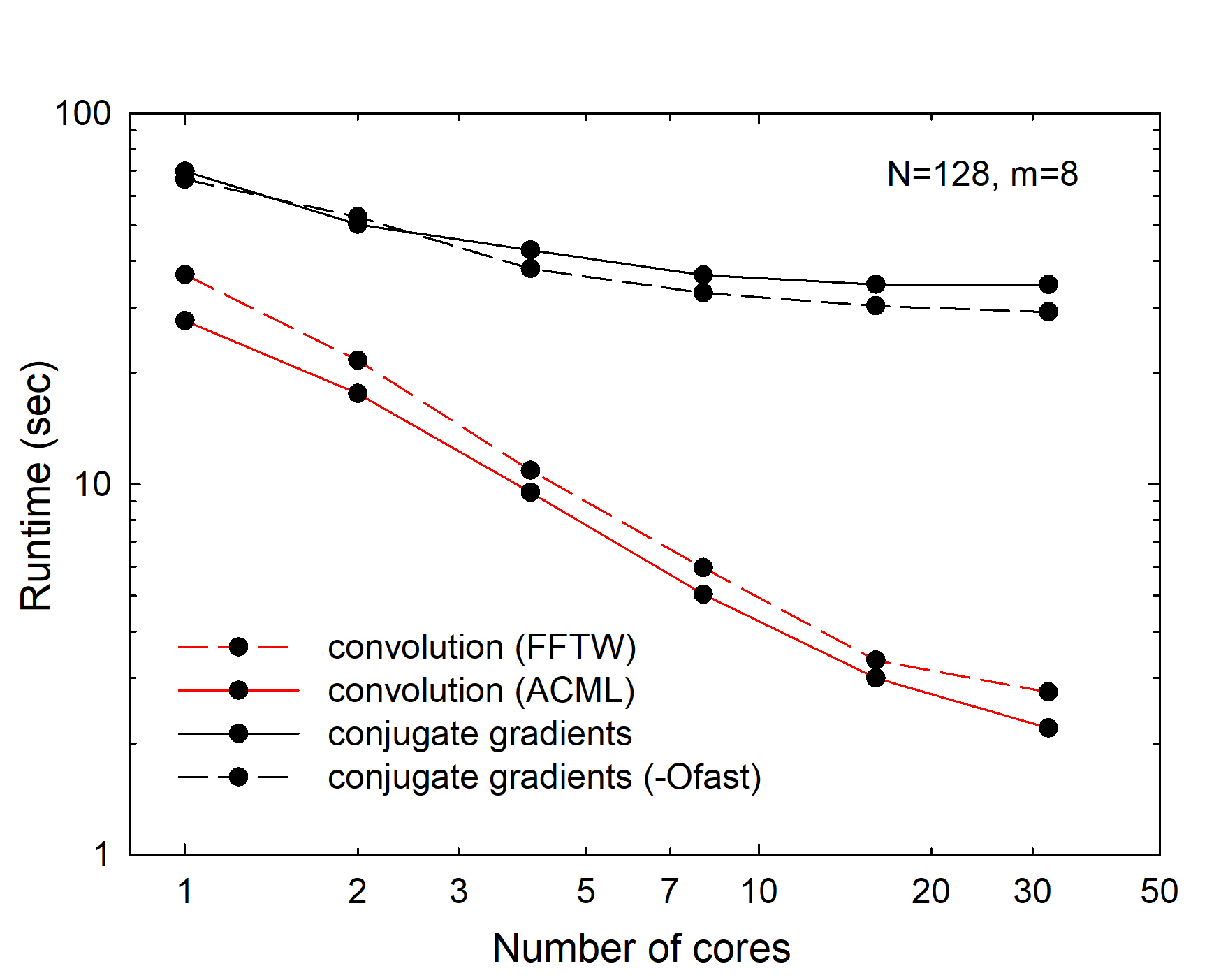}
\caption{Runtime vs. the number of cores in computing the gravitational potential of an oblate ellipsoid on $m=8$ nested grids. Red and black lines correspond to the convolution method and OiCG method, respectively. 
}
\label{fig:13}
\end{figure}

\section{Performance comparison}
\label{Sect:performance}
In this section, we compare the two methods for computing the gravitational potential in terms of its performance on the Intel Xeon Platinum 8174 processors with 48 physical cores. We run the test in a loop that has 20 full cycles of the potential calculation in order to reduce the associated overheads and then divide the obtained runtime by a factor of 20. Figure~\ref{fig:13} presents the results of our tests for the number of OpenMP threads ranging from 1 to 32, increasing by a factor of 2. The basic tests use the Intel/19 FORTRAN compiler without explicit optimization. Additionally, we run tests with the -Ofast optimization key. For the test configuration, we have chosen an oblate ellipsoid as in Sect.~\ref{Compare}. The resolution is $N=128$ cells in each direction and the number of nested grids is $m=8$. Both values are based on our practical simulations of cloud core collapse and protoplanetary disk formation in the Solar metallicity environment, resulting in a numerical resolution of $\approx 1$~au near the forming star. The tolerance $\epsilon$ for the OiCG method is set equal to $10^{-8}$. To compute the FFTs, we used two external libraries: the AMD Core Math Library (ACML) and the FFTW\footnote{\href{https://fftw.org}{https://fftw.org}} library. 

Analysis of Fig.~\ref{fig:13} indicates that the convolution method outperforms the OiCG method by more than an order of magnitude on multicore OpenMP applications. Even without parallelization, the convolution method is faster than the OiCG method by a factor of several. The optimization key -Ofast marginally improves the performance of the OiCG method but has little effect on the convolution method. The ACML library is found to be somewhat faster than that of FFTW.

\begin{figure}
\centering
\includegraphics[width=1\columnwidth]{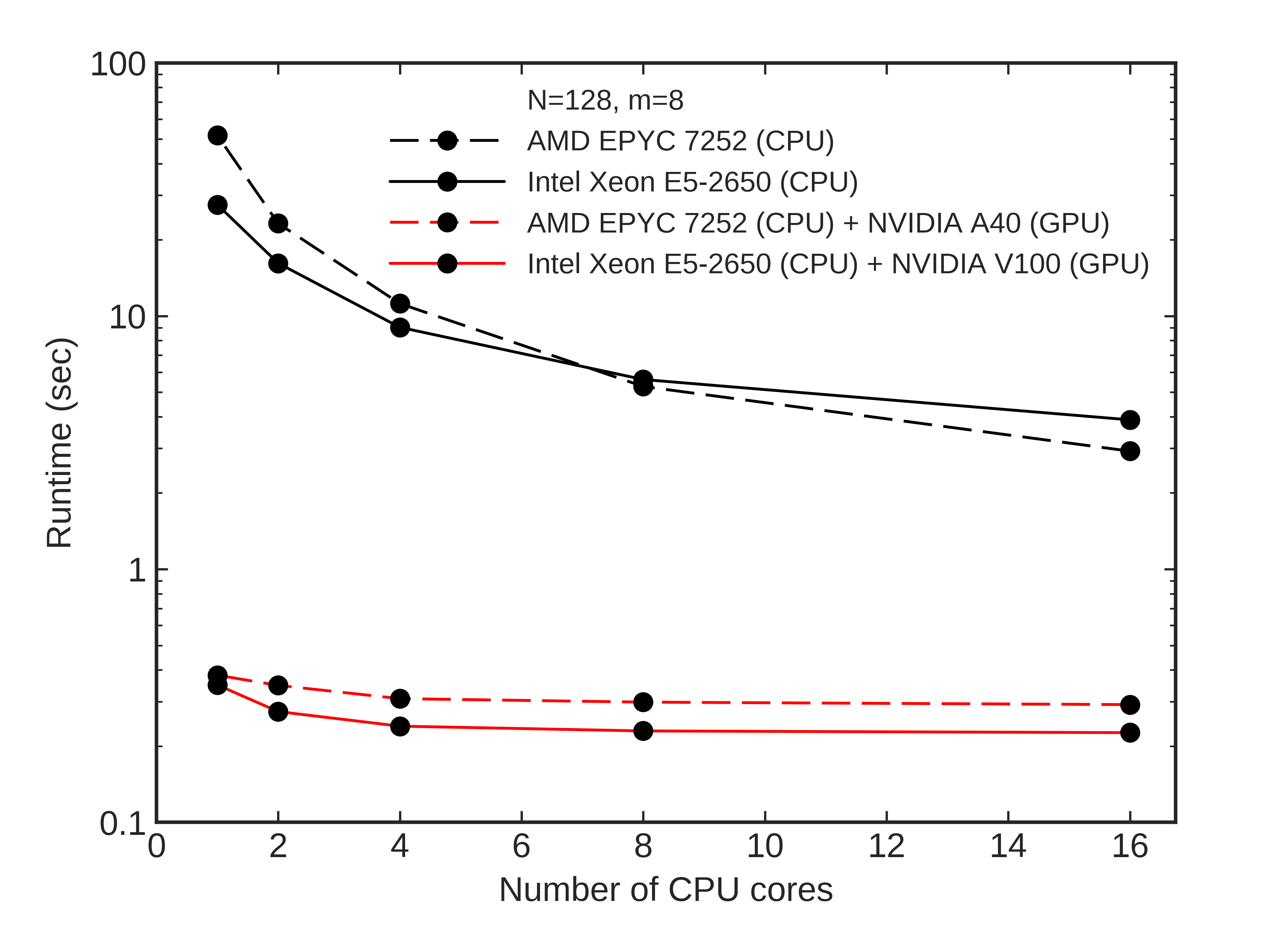}
\caption{Runtime vs. the number of CPU cores in computing the gravitational potential of an oblate ellipsoid on $m=8$ nested grids. The red lines show the results of CPU plus GPU computations, while the black lines present the CPU model only.
}
\label{fig:14}
\end{figure}

An even better performance on the convolution method can be achieved when graphics processing units (GPUs) are utilized. While an extension of the Intel FORTRAN compiler can offer an interface with GPUs (hereafter, devices) through OpenMP target offload, CUDA by NVIDIA is widely accepted as faster in most cases and thus was the chosen method of implementation \citep{Gayatri2019}. This required the adoption of the NVIDIA FORTRAN compiler\footnote{\href{https://developer.nvidia.com/cuda-fortran}{https://developer.nvidia.com/cuda-fortran}}. Various NVIDIA devices were used to accelerate the computations of the gravitational potential on each nested mesh, and this was coupled with CPU (hereafter, host) calculations of the boundary conditions and the final assembly of the gravitational potential from potentials on each individual mesh (see eq.~\ref{eq:meshes}). The results presented here are for the NVIDIA A40 and NVIDIA V100, the parameters for which can be found in Table \ref{tab:device_configs}. While the host of the NVIDIA V100, the Intel Xeon E5-2650 v3, offers a maximum of 20 physical cores, for consistency and in keeping with the dimensions of the computational domain, results are presented to a maximum of 16 physical cores. To compute the FFTs, the CUDA library \textit{cufft} and the \textit{iso\_c\_binding} module for FORTRAN to C interoperability were used. As in the example shown in Figure~\ref{fig:13}, an oblate ellipsoid was chosen, with a numerical resolution of $N=128$ grid zones in each coordinate direction and $m=8$ nested meshes. The results are presented in Figure~\ref{fig:14}. The CPU-GPU coupled version can be seen to achieve acceleration by a factor of 10 compared to the pure CPU model when at full OpenMP host parallelization. During operations accelerated by the device, 75\% of the time is spent performing FFT computation, 20\% on transferring the data between the host and the device, and 5\% on other auxiliary computations. In total when using GPUs, a factor of 200 times acceleration can be achieved when compared to the original OiCG method with CPU-only implementation (see the black lines in Fig.~\ref{fig:13}).

\begin{table*}[]
    \centering
    \caption{Different device-host configurations}
    \begin{tabular}{llll|ll}
    \hline
    \hline
    \multicolumn{4}{c|}{Device} & \multicolumn{2}{c}{Host} \\
    Name & \addstackgap{\shortstack[l]{CUDA\\cores}} & \addstackgap{\shortstack[l]{GFLOPs/s\\(FP64)}} & \addstackgap{\shortstack[l]{CUDA\\version}} & Name & \addstackgap{\shortstack[l]{Physical\\cores}}\\ \hline
    \addstackgap{\shortstack[l]{NVIDIA\\A40}} & 10752 & 1169 & 11.6 & \addstackgap{\shortstack[l]{AMD EPYC\\7252 (x2)}} & 16 \\
    \addstackgap{\shortstack[l]{NVIDIA\\V100}} & 5120 & 7000 & 11.2 & \addstackgap{\shortstack[l]{Intel Xeon\\E5-2650 v3 (x2)}} & 20
    \end{tabular}
    \label{tab:device_configs}
\end{table*}

Finally, we tested the performance of the Generalized Minimal Residual method (GMRes) when computing the gravitational potential using the discretized form of the Poisson equation (see, eqs.~\ref{poisson:eq} and \ref{poisson:descrete}). In our particular case, we applied the OiCG method outlined in Sect.~\ref{inside-out} to find the potential on each grid. Intel/19 FORTRAN compiler and the MKL Intel library were used for this test problem. No explicit optimization was used. We did this additional test to illustrate how our method can outperform the ``standard'' methods offered by frequently used numerical libraries. GMRes is an iterative method and, similar to the conjugate gradient method, it requires presetting a desired accuracy when calculating the gravitational potential. We set the tolerance in the relative residuals to $\epsilon=10^{-6}$ to be consistent with the OiCG method. 
We found, however, that GMRes does not converge on $N=128$ grid zones unless we reduce the tolerance $\epsilon$. Therefore, we decreased the number of grid zones in each direction to $N=64$. GMRes also required too many iterations to converge for nested grid levels $m \ge 7$, which may be related to the problem of divergence discussed in the context of Figure~\ref{fig:12}. We, therefore, limited our test runs to $m=6$. The runtime for the GMRes method for an oblate ellipsoid (again, using 20 loops of full potential calculations to reduce the overheads) is 29 seconds and this number weakly depends on the OpenMP parallelization. This value is comparable to the best performance of the OiCG method but is much worse than the performance of the convolution method. Considering that the test was done on a twice lower resolution and with a smaller number of nested grids, the GMRes methods of the Intel MKL library is clearly inferior to both methods presented in this work. The GMRes method can be accelerated by reducing the tolerance $\epsilon$, but we found that this leads to degraded accuracy in the calculation of the potential on the innermost nested grid and is therefore not recommended.

\section{The dipole problem and the modified convolution method}
\label{sect:dipole}
\begin{figure}
\centering
\includegraphics[width=1\columnwidth]{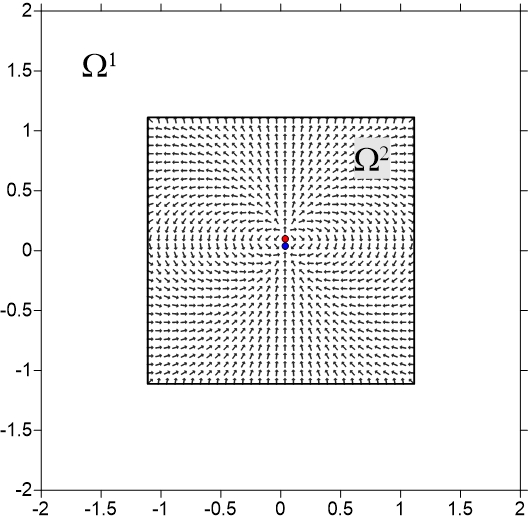}
\caption{ Electric field vectors of a dipole schematically shown by the red and blue circles near the coordinate center. The original convolution method is applied to construct the field. We note that the field is present and regular on the fine $\Omega^2$ grid but is absent on the coarse $\Omega^1$ grid. This occurs because of the cancellation of the charges when they are projected from the fine to the coarse grid in the original convolution method. 
}
\label{fig:15a}
\end{figure}

The original convolution method described in Sects.~\ref{convolution} and \ref{nested} can be modified if specific cases are encountered. We illustrate this using the following idealized example problem. We consider an electric dipole with a dipole moment $p=q d$, where $d$ is the distance between unit charges $+q$ and $-q$. For simplicity, we place the dipole on a nested mesh with only two nested grids $\Omega^1$ and $\Omega^2$. Each grid has a numerical resolution of $N=32^3$. The unit charges are located near the geometrical center of the fine $\Omega^2$ grid and their spatial position is chosen so as to cancel each other on the course $\Omega^1$ grid. This means that on the coarse grid, the net charge is zero. The resulting electric field geometry is illustrated in Figure~\ref{fig:15a}. While the field lines show a regular and correct pattern on the fine $\Omega^2$ grid, the field is artificially suppressed on the course $\Omega^1$ grid because of the charge cancellation. The problem was outlined and successfully addressed in \citet{2007Matsumoto, 1998Truelove, 2020Stone}.

\begin{figure}
\centering
\includegraphics[width=1\columnwidth]{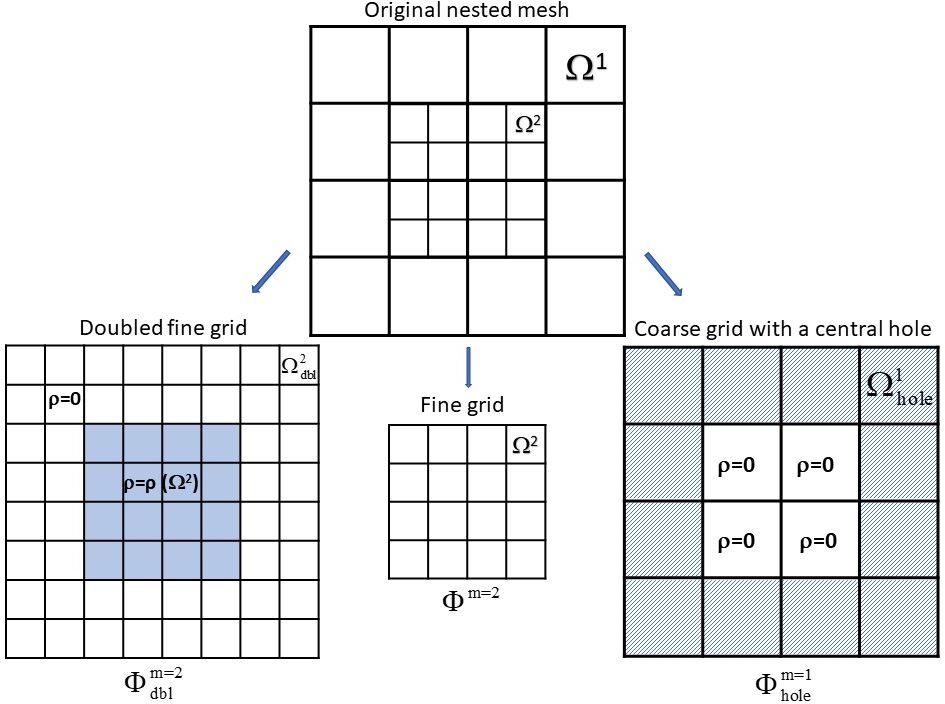}
\caption{Decomposition of two nested grids (top) into three individual grids in the modified convolution method: a fine $m=2$ grid ($\Omega^2$, bottom center), a coarse $m=1$ grid with a central cavity ($\Omega^1_{\rm hole}$, bottom right), and a doubled fine grid $m=2$ ($\Omega^2_{\rm dbl}$, bottom left). The density in the latter grid is set to zero everywhere except the original physical size of the fine $m=2$ grid (shown with a light blue shade), where the density inherits the values from the fine m=2 grid, $\rho=\rho(\Omega^2)$.
}
\label{fig:16}
\end{figure}

Our augmentation of the convolution method is described in detail below.
We do not consider possible updates to the OiCG method as this method is used mainly for comparison purposes. Figure~\ref{fig:16} presents an example of the mesh decomposition, which can be used to solve the aforementioned dipole problem on two nested grids. In particular, the solution procedure is split into three steps. The first two steps are identical to the original method: the potential on a fine $m=2$ grid ($\Omega^2$, bottom center) and on a coarse $m=1$ grid with a central cavity ($\Omega^1_{\rm hole}$, bottom right) is found. The third step, however, involves expanding the fine grid to the physical size of the coarse grid as shown in the bottom left panel. The resulting doubled fine grid $\Omega^2_{\rm dpb}$ has the same physical size as the coarse grid $\Omega^1$ but the size of individual numerical cells is similar to that of $\Omega^2$. As a result, $\Omega^2_{\rm dpb}$ has twice as many cells in each coordinate direction. The additional layers of cells are then filled with zeros (white cells in the bottom left panel), while the original physical size of the fine $m=2$ grid (indicated by the light-blue shade) retains the original values from the fine $\Omega^2$ grid. The described algorithm can be written as follows
\begin{eqnarray}
\label{eq:meshes2}
    \Phi(\Omega^1) &=&  \Phi^{\rm m=1}_{\rm hole}+ \tilde{\Phi}^{\rm m=2}_{\rm dbl}, \nonumber \\
    \Phi(\Omega^2) &=&  \tilde{\Phi}^{\rm m=1}_{\rm hole}+ \Phi^{\rm m=2}.
\end{eqnarray}

\begin{figure}
\centering
\includegraphics[scale=0.4]{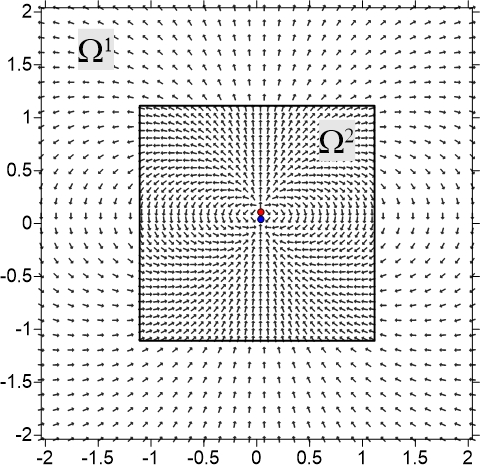}
\includegraphics[scale=0.4]{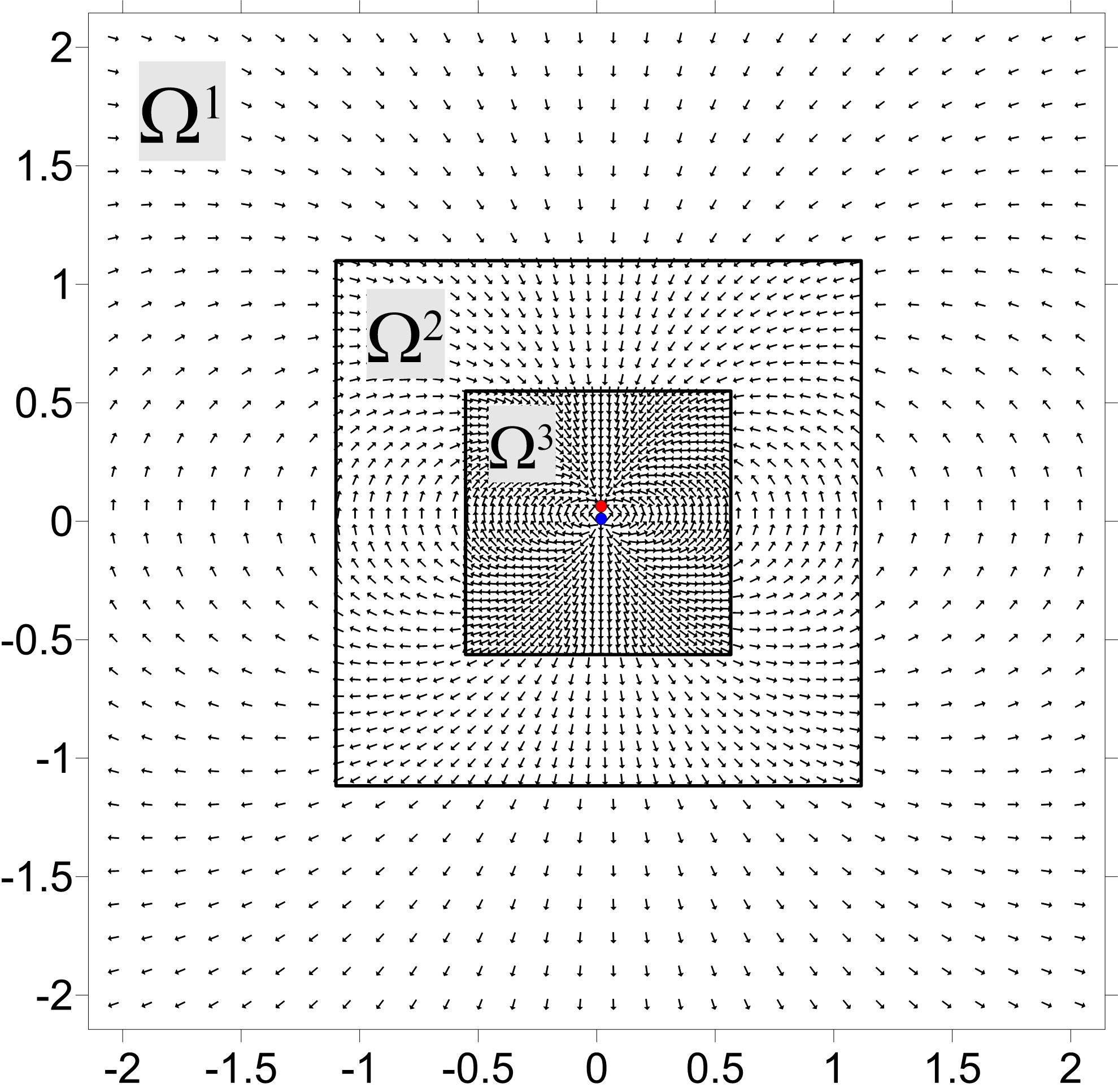}
\caption{ Electric field vectors of a dipole. The dipole is located at the center of the finest grid and is shown schematically by the red and blue circles. The modified convolution method is applied to construct the field (see Fig.~\ref{fig:15a} for comparison with the original method). The top panel presents the case of two nested grids, while the bottom panel corresponds to three nested grids. The black squares identify the grid interfaces. 
Note that the field is present and continuous across all nested grids $\Omega^1$, $\Omega^2$ and $\Omega^3$. 
}
\label{fig:17}
\end{figure}

The introduction of the doubled fine grid allows us to avoid the charge cancellation on the coarse grid and retrieve the correct behavior of the electric field lines across the grid interface as illustrated in the top panel of Figure~\ref{fig:17}. Now, the field lines are continuous and are present on the entire nested mesh (and not only on the fine mesh as was in the original method). 
Finding the potential on the coarse grid $\Omega^1$ is now a more complex procedure than it was in the original method (see Eq.~\ref{eq:meshes}). The potential on the doubled fine grid $\Phi^{\rm m=2}_{\rm dbl}$ should further be reduced to the coarse $m=1$ grid (this mathematical operation is denoted by tilde) and added to the potential on the coarse grid with a hole $\Phi^{\rm m=1}_{\rm hole}$. We note that the above procedure can in principle be extended to any number of nested grids (see Appendix~\ref{sect:App0}). For instance, the bottom panel in Figure~\ref{fig:17} presents the electric field of a dipole on three nested grids. As can be seen, the field lines are present and smooth across all grids.
However, the modified convolution method suffers from computational overheads and we discuss this issue in more detail in Appendix~\ref{sect:App0}.

We want to make the following notes in this context. The electric dipole problem considered above is an extreme case, which helped us to highlight the problem. A close analog of an electric dipole in astrophysics is a binary star. However, masses are never cancelled out but add up to produce the common gravitational field. We have already considered a wide-separation binary system in Sect.~\ref{convolution} and our original convolution method has revealed good performance on this nonaxisymmetric test problem (see also Appendix~\ref{App:compare}), but the situation with close binaries may be different. The dipole component of the gravity force of a close binary star falls off with distance faster ($\sim 1/r^3$) than the basic field component ($\sim 1/r^2$). This means that special care should be taken when computing the potential of a binary star in its vicinity, where the dipole moments are important, but at distances much greater than the binary separation the dipole contribution may be insignificant.

\begin{figure}
\centering
\includegraphics[width=1\columnwidth]{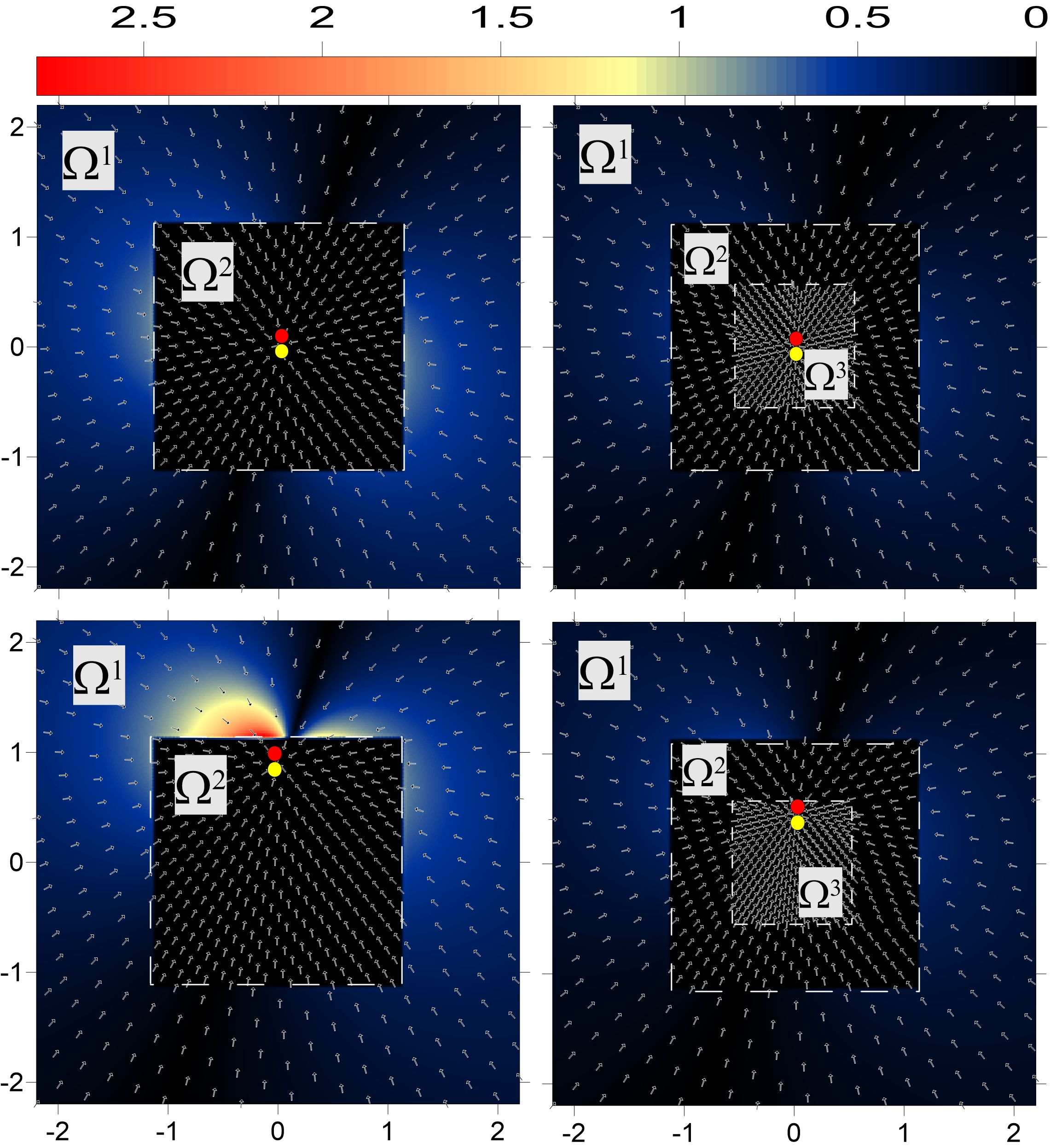}
\caption{ Relative errors between the gravitational potentials of a close binary computed by the original and modified convolution methods. The vertical bar shows the errors in percent. Two mesh setups are used for this test problem: left column -- two nested grids and right column -- three nested grids. The grid interfaces are shown by the dashed lines for convenience. The binary position is schematically indicated by the red and yellow circles. The arrows show the gravitational field in the modified convolution method. Note that the field lines are continuous across the grid interfaces. 
}
\label{fig:18}
\end{figure}

We illustrate the problem in Figure~\ref{fig:18} by computing the gravitational field of a close binary in the original and modified convolution methods. We consider a tight binary with companions located next to each other on the finest grid ($\Omega^2$ for two nested grids and $\Omega^3$ for three nested grids). The densities in the neighboring cells occupied by the binary are set to $\rho=1$ and $\rho=2$, implying the companion mass ratio 1:2, and are zeroed everywhere else in the computational domain. We then calculate the relative error (in per cent) between the two methods, keeping in mind that the modified method computes the dipole moments accurately, but the original method does not. The top panel corresponds to the binary placed in the center of the finest grid. The bottom panel presents the case when the binary is located near the upper interface between $\Omega^1$ and $\Omega^2$ when two nested grids are considered and near the interface of $\Omega^2$ and $\Omega^3$ for the case of three nested grids. Each level of refinement has $N=128^3$ grid cells.

We first consider two nested grids shown in the left column of Figure~\ref{fig:18}.
Both methods produce identical solutions on the refined $\Omega^2$ grid and, hence, the relative error is zero there. The deviations between the two methods occur on the coarse $\Omega^1$ grid, for which the original method does not reproduce correctly the dipole moments, and the errors depend on the spatial location of the binary. The errors are highest when the binary is near the grid interface (bottom panel) and can reach several per cent on the coarse grid. On the other hand, the errors stay below 1\% if the binary is in the center of the fine grid (top panel). 
The case of three nested grids is shown in the right column of Figure~\ref{fig:18}. Here, the comparison is made between the modified convolution method that correctly computes the dipole moments only on the finer $\Omega^2$ and $\Omega^3$ grids, but not on the coarsest $\Omega^1$ grid, and the modified convolution method that correctly computes the dipole moments on all three nested grids. Both cases produce identical potentials in the $\Omega^2$ and $\Omega^3$ grids and the relative error is zero there, but they deviate in the outermost $\Omega^1$ grid. However, the errors stay low and do not exceed 0.35\%, regardless of the spatial position of the binary. If we degrade the numerical resolution to $N=64^3$ and $N=32^3$, the maximum errors slightly increase to 0.7\% and 1.5\%, which can be expected with the overall decrease in accuracy of the method as the resolution drops.  

These test examples demonstrate the utility of the updated convolution method in correctly reproducing the dipole field of a binary across the coarse-fine grid interfaces. But it also shows that the error quickly drops with distance from the binary. A similar behavior was found when we varied the separation between the binary components by several grid cells. Consideration between the associated numerical costs and desired accuracy is needed if the dipole moments over a larger number of grid interfaces are to be computed. For instance, the convolution method was originally developed for astrophysical problems involving the gravitational collapse of an individual prestellar core. The star(s) and the circumstellar disk that form in the process are located within two-three innermost grids and the rest is filled with a rarified infalling envelope. The correct calculation of the dipole moments over the three innermost nested grids should be sufficient in this case. We also note that when applied to a smooth density distribution, e.g., an oblate ellipsoid, or wide separation binaries we did not find significant differences between the original and modified convolution methods (see Appendix~\ref{App:compare}) for details).

\begin{figure}
\centering
\includegraphics[width=1\columnwidth]{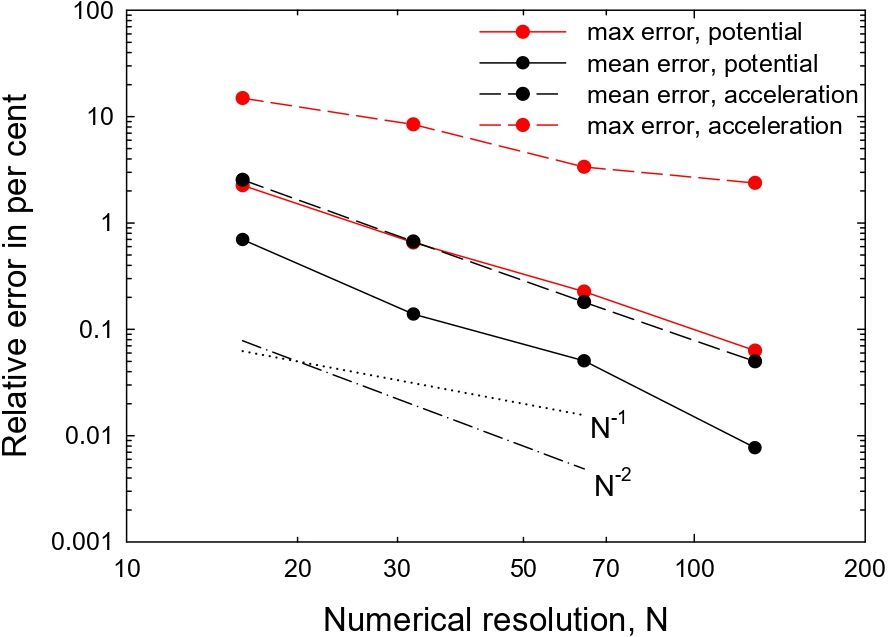}
\caption{ Convergence test in the modified convolution method. Shown are the mean (black lines) and maximum (red lines) errors in the gravitational potential and acceleration as a function of numerical resolution $N$. The dotted and dash-dotted lines show the linear and quadratic convergence for convenience.
}
\label{fig:20}
\end{figure}

Finally, we performed a convergence test on the modified convolution method, in which we varied the resolution of each nested grid from $N=16$ to $N=128$, but kept the number of nested grid fixed at $m=4$. An oblate ellipsoid with the ratio of semi-axes 0.5:1.0:1.0 was used for this purpose.
The results are shown in Figure~\ref{fig:20}. When the mean numerical errors are considered, the method demonstrates second-order accuracy in both the gravitational potential and acceleration. The maximum error is characterized by second-order convergence only for the gravitational potential, while the maximum error in acceleration reveals first-order convergence. 
We attribute this decrease in accuracy to boundary effects between the nested grids, where the adopted linear interpolation procedure of the gravitational potential is only first-order accurate. The development of more accurate interpolation schemes across the grid interfaces should be our next task in future works.

\section{Conclusions}
\label{conclude}

In this paper, we presented an extension of the convolution method for calculating the gravitational potential on regular and equidistant grids, which can be applied to the practical case of nested grids. We compared the performance of the convolution method against the OiCG method using two test problems: an oblate ellipsoid and a binary star, for which the analytic solution of the gravitational potential and acceleration are known. We found that both methods are characterized by comparable computational errors, which do not exceed a few percent for the gravitational acceleration and are notably lower for the gravitational potential. 

The convolution method, however, has several clear advantages over the iterative OiCG method. First, the convolution method does not require the precalculation of the gravitational potential at the boundaries of the coarsest $m=1$ grid. Second, the convolution method permits decomposition of the nested grids into individual subgrids, followed by efficient computation of the potential on each subgrid in a parallel mode. Third, the convolution method does not require a preset tolerance $\epsilon$, with which the iterations in the OiCG method are controlled. We found that the OiCG method reveals an increase in the errors with increasing $m$, and the magnitude of this effect depends on the predefined value of $\epsilon$.  

A comparison of runtimes of the OiCG and convolution methods demonstrates a clear advantage of the latter method, which can outperform the former by more than a factor of 10 on the CPUs with OpenMP parallelization. When GPUs are utilized, the convolution method can be accelerated by another factor of 10. The GMRes method offered by the Intel MKL library is found to be inferior in speed performance as compared to both methods presented in this work. The convolution method revealed an overall second-order convergence, except for the maximum error at the coarse-fine grid interfaces where the convergence is linear, in accordance with the adopted linear interpolation of the gravitational potential across the grid interface.

The disadvantage of the convolution method lies in the requirement to interpolate the gravitational potential from the coarse grids onto the refined grids. This step can be improved by devising more sophisticated interpolation procedures. When tightly spaced gravitating sources are present at the deepest levels of refinement, for example, close binary/multiple stars surrounded by a circumstellar disk, a modification to the convolution method is required to correctly calculate the dipole moments on the coarser grids. The modification introduces computational overheads, which may be substantial for a large number of nested grids, but 
acceptable accuracy can already be reached when the method is applied to each pair or triplet of multiple nested grids.

\section{Acknowledgements}
We are thankful to the referee for the comments and suggestions that helped to improve the manuscript. This work was supported by the FWF project I4311-N27 (E. I. V. and J. M.) and RFBR project 19-51-14002 (I. K. and V. E.). The simulations were performed on the VSC Vienna Scientific Cluster. We thank Oliver Hahn for pointing to an analytic solution of Eq.~(\ref{pot:cell}). The code that uses the convolution theorem for calculating the gravitational potential on CPUs and GPUs can be employed  on reasonable request.




\bibliographystyle{aa}
\bibliography{refs}

\begin{appendix}

\section{Gravitational acceleration at the coarse-fine grid interface}
\label{sect:App1}

\begin{figure}
\centering
\includegraphics[width=1\columnwidth]{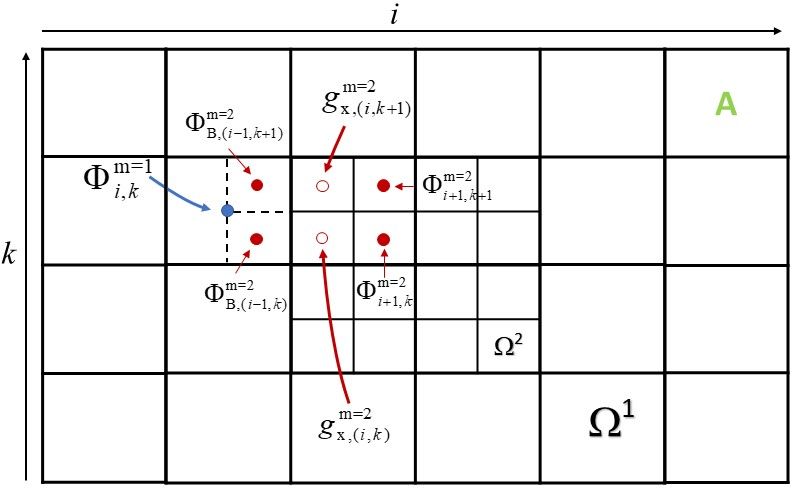}
\includegraphics[width=1\columnwidth]{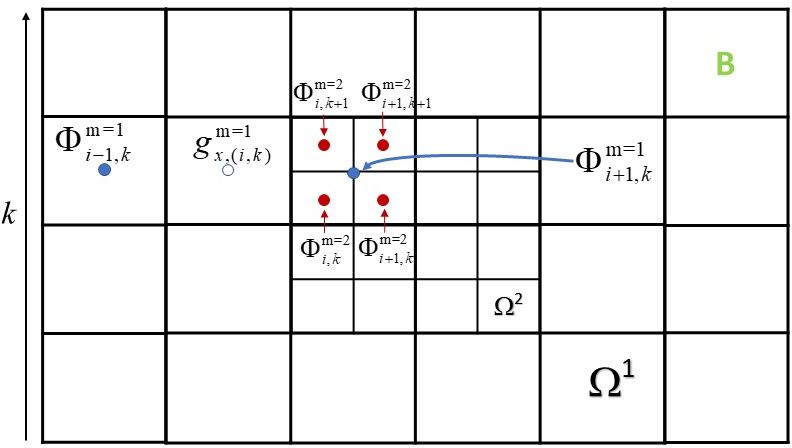}
\includegraphics[width=1\columnwidth]{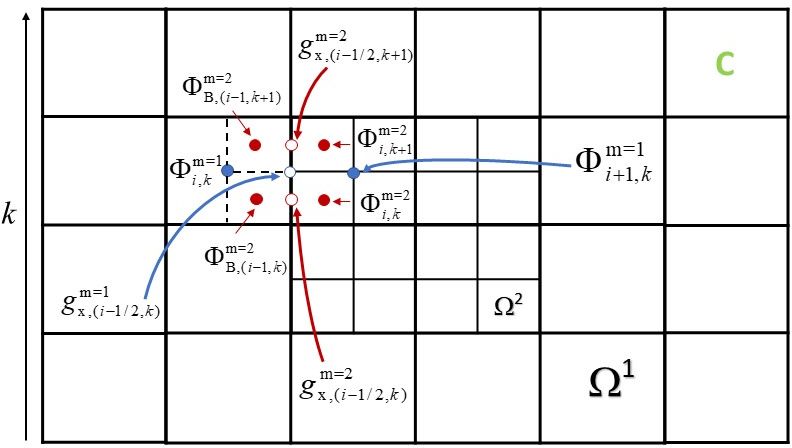}
\caption{Three methods to calculate the gravitational acceleration at the cells adjacent to the coarse-fine grid interface. {\bf Top panel.} Method A schematically illustrates a scheme to calculate the acceleration at the adjacent fine cells. {\bf Middle panel}. Method B presents a scheme to calculate the acceleration at the adjacent coarse cells. {\bf Bottom panel} Method C shows yet another scheme to calculate the acceleration at the adjacent coarse cells, which fulfills the Gauss theorem (see the text for details).
}
\label{fig:App1}
\end{figure}

The components of gravitational acceleration ${\bl g}$ across the coarse-fine grid interface can be calculated using three methods illustrated in Figure~\ref{fig:App1}. We take two nested grids, $\Omega^1$ and $\Omega^2$, for simplicity and consider two dimensions, with the indices $i$ and $k$ corresponding to the $x$- and $y$-coordinates, respectively. The schemes can straightforwardly be extended to three dimensions.

Panel A graphically describes the case of the fine grid cells that are adjacent to the coarse grid (hereafter, method A). The $x$-component of the gravitational acceleration at the fine-to-coarse grid interface is calculated as
\begin{equation}
    g^{m=2}_{{\rm x},(i,k)} = - {\Phi^{\rm m=2}_{i+1,k} - \Phi^{\rm m=2}_{{\rm B},(i-1,k)} \over 2\,  dx_{\rm m=2} },
\end{equation}
where $\Phi^{\rm m=2}_{{\rm B},(i-1,k)}$ are the boundary values of the potential on the fine grid found using the interpolation scheme described in Fig.~\ref{fig:2} of Sect.~\ref{inside-out} and $dx_{\rm m=2}$ is the size of the fine grid cell.

Panel B presents the opposite case of the coarse grid cells that are adjacent to the fine grid (hereafter, method B). The gravitational acceleration at the coarse-to-fine grid interface can be calculated as
\begin{equation}
    g^{m=1}_{{\rm x},(i,k)} = - {\Phi^{\rm m=1}_{i+1,k} - \Phi^{\rm m=1}_{(i-1,k)} \over 2\,  dx_{\rm m=1} },
    \label{coarse-to-fine}
\end{equation}
where $\Phi^{\rm m=1}_{(i-1,k)}$ is found using an arithmetic average of $\Phi^{\rm m=2}_{(i,k)}$, $\Phi^{\rm m=2}_{(i+1,k)}$, $\Phi^{\rm m=2}_{(i,k+1)}$, and $\Phi^{\rm m=2}_{(i+1,k+1)}$. Here, $dx_{\rm m=1}$ is the size of the coarse grid cells.

Panel C presents a more subtle and physically motivated method (hereafter, method C) for calculating the gravitational acceleration at the coarse-to-fine grid interface \citep[see also][]{2018Feng}. Here, we first calculate the $x$-components of gravitational acceleration at the coarse-fine grid interface using the coarse and fine grid values separately
\begin{eqnarray}
    g^{m=1}_{{\rm x},(i-1/2,k)} &=& - {\Phi^{\rm m=1}_{i+1,k} - \Phi^{\rm m=1}_{(i,k)} \over dx_{\rm m=1} },  \label{eq:grav_coarse0} \\
    g^{m=2}_{{\rm x},(i-1/2,k)} &=& - {\Phi^{\rm m=2}_{i,k} - \Phi^{\rm m=2}_{{\rm B},(i-1,k)} \over dx_{\rm m=2} }, \\
    g^{m=2}_{{\rm x},(i-1/2,k+1)} &=& - {\Phi^{\rm m=2}_{i,k+1} - \Phi^{\rm m=2}_{{\rm B},(i-1,k+1)} \over dx_{\rm m=2} }.
    \label{eq:grav_coarse}
\end{eqnarray}
We note that these values of gravitational acceleration are defined at the cell interfaces and not at the cell centers. We then require that the gravitational accelerations at the coarse-fine grid interface be equal when using the coarse and fine grids separately 
\begin{equation}
    g^{m=1}_{{\rm x},(i-1/2,k)} = {1\over 2}(g^{m=2}_{{\rm x},(i-1/2,k)} + g^{m=2}_{{\rm x},(i-1/2,k+1)}).
    \label{coarse-to-fine_2}
\end{equation}
From the last expression it is straightforward to find the value of $\Phi^{\rm m=1}_{i+1,k}$ and then employ Equation~(\ref{coarse-to-fine}) to compute the gravitational acceleration $g^{m=1}_{{\rm x},(i,k)}$ at the coarse grid cells that are adjacent to the fine grid. This scheme is not fully self-consistent in the sense that the value of $\Phi^{\rm m=1}_{i+1,k}$ depends on the boundary values on the fine grid $\Phi^{\rm m=2}_{{\rm B},(i-1,k)}$ and $\Phi^{\rm m=2}_{{\rm B},(i-1,k+1)}$, which in turn are found using the potentials on the coarse grid (see Eqs.~\ref{eq:bound1}--\ref{eq:bound2}). This inconsistency can be resolved through iterations when the value of $\Phi^{\rm m=1}_{i+1,k}$ is substituted back to Eqs.~(\ref{eq:bound1})--(\ref{eq:bound2}) to find the boundary potentials $\Phi^{\rm m=2}_{{\rm B},(i-1,k)}$ and $\Phi^{\rm m=2}_{{\rm B},(i-1,k+1)}$ and the calculations involving Eqs.~(\ref{eq:grav_coarse0})--(\ref{coarse-to-fine_2}) are repeated. Our experience shows that convergence is achieved after several iterations.

\begin{figure}
\centering
\includegraphics[width=1\columnwidth]{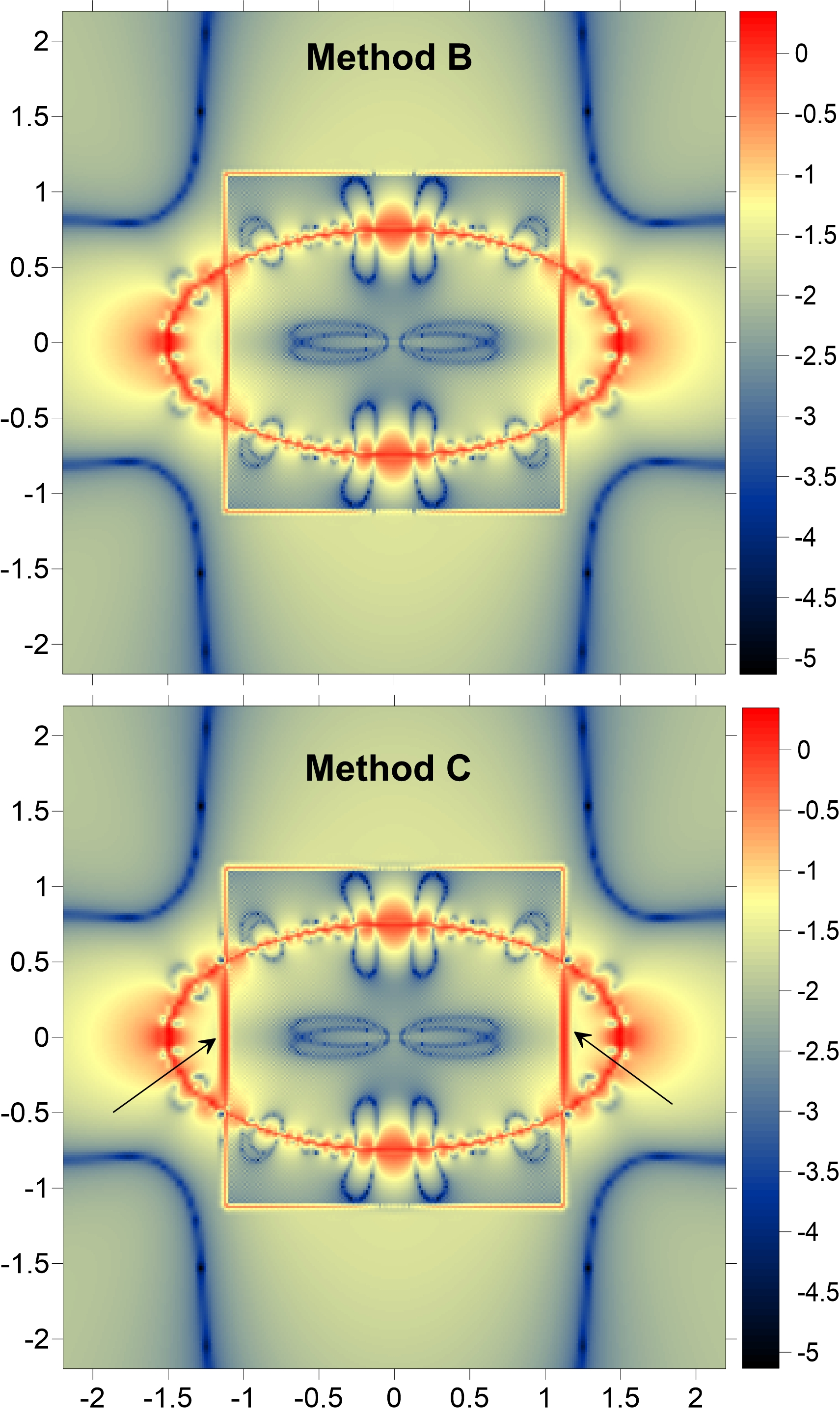}
\caption{ Relative error in the gravitational acceleration of an oblate ellipsoid. The top and bottom panels correspond to methods B and C for computing the gravitational acceleration at the coarse cells that are adjacent to the coarse-fine grid interface. The arrows illustrate the regions where differences between methods B and C are found. The color bars show errors relative to the analytic solution in per cent.
}
\label{fig:App2}
\end{figure}

The difference between methods B and C is illustrated in Figure~\ref{fig:App2}, which shows the relative error in the gravitational acceleration at the $x-z$ plane obtained for both methods. We consider two nested grids and an ellipsoid with the ratio of semi-axes 0.75:1.5:1.5. We note that for both cases, method A is used to calculate the acceleration in the cells of the fine grid that are adjacent to the coarse grid. We found that the span of errors is similar in both methods. In particular, the maximum errors occur in the vicinity of the ellipsoid surface and also at the coarse-fine grid interface. When method B is considered, the error is maximal at the adjacent fine cells and is much smaller at the adjacent coarse cells. When method C is considered, the error is similar and maximal on both sides of the coarse-fine grid interface. This effect is noticeable if the coarse-fine grid interface lies inside the ellipsoid, as illustrated by the arrows in Figure~\ref{fig:App2}. The corresponding region with the highest error appears thicker in method C than in method B. To summarize, method C is physically more accurate as it fulfills the Gauss theorem at the grid interfaces, but it comes at the expense of an increased error in the gravitational acceleration at the coarse cells that are adjacent to the fine grid.

\section{Modified convolution method and reduction of overheads} 
\label{sect:App0}

In this section, we describe how the convolution method can be extended to tackle the dipole problem across multiple nested grids and highlight the associated numerical costs. First, we provide a scheme, in which the dipole moments are accurately computed for each pair of neighboring grids on a nested mesh with $m\ge 3$. As was demonstrated in Sect.~\ref{sect:dipole}, this allows notably reducing the error in the gravitational potential of a close binary near the coarse-fine grid interface. The corresponding algorithm can be written as
\begin{eqnarray}
\label{App:first}
     \Phi(\Omega^n) &=& \Phi^{\rm m=n}_{\rm hole} +\tilde{\Phi}^{\rm m=n+1}_{\rm dbl}+ \sum_{k=1}^{n-1} \tilde{\Phi}^{\rm m=n}_{\rm hole}, \,\, \mathrm{for} \,\, n<M_{\rm depth}   \\   
     \Phi(\Omega^n) &=& \sum_{k=1}^{n-1} \tilde{\Phi}^{\rm m=n}_{\rm hole} + \Phi^{\rm m=n}, \,\, \mathrm{for} \,\, n=M_{\rm depth}.
     \label{App:first2}
\end{eqnarray}
Here, $M_{\rm depth}$ is the maximum number of nested grids. Each pair of neighboring grids $\Omega^n$:$\Omega^{n+1}$ has now its own auxiliary doubled grid $\tilde{\Phi}^{\rm m=n+1}_{\rm dbl}$. As usual, the tilde sign means proper interpolation from a coarser to a finer grid and vice versa.
When this modified scheme is used on CPUs, the runtime increases by about a factor of four compared to the original method. When CPUs and GPUs are used together, the runtime increases by a factor of 2.

\begin{figure}
\centering
\includegraphics[width=1\columnwidth]{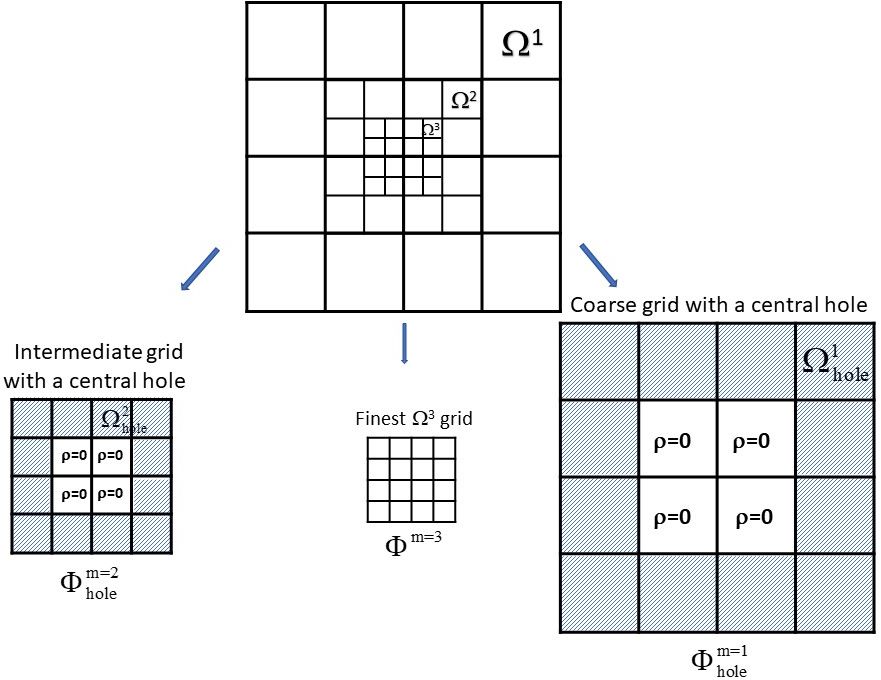}
\includegraphics[width=1\columnwidth]{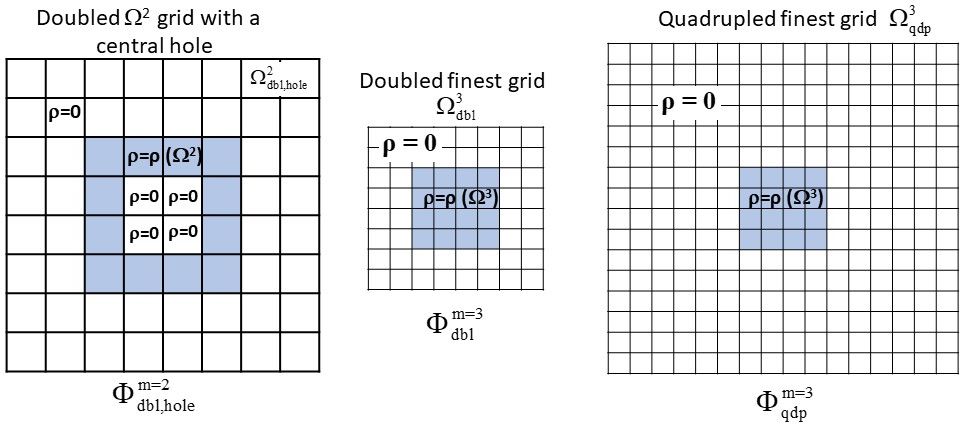}
\caption{Decomposition of the $m=3$ mesh in the modified convolution method accounting for the dipole moments. Six individual grids are now considered with the three grids in the bottom row representing the auxiliary steps needed to compute the dipole moments for three neighboring grids.
}
\label{fig:App0}
\end{figure}

When the dipole moment are to be computed across three neighboring grids (as in the bottom panel of Fig.~\ref{fig:17}), the corresponding algorithm reads as
\begin{eqnarray}
\label{eq:meshes3}
    \Phi(\Omega^n) &=& \Phi^{\rm m=n}_{\rm hole} +\tilde{\Phi}^{\rm m=n+1}_{\rm dbl,hole}+ \tilde{\Phi}^{\rm m=n+2}_{\rm qdp} + \sum_{k=1}^{n-1} \tilde{\Phi}^{\rm m=n}_{\rm hole}, \,\, \\
            &\,& \mathrm{for} \,\, n<M_{\rm depth}-1   \\
     \Phi(\Omega^n) &=& \Phi^{\rm m=n}_{\rm hole} +\tilde{\Phi}^{\rm m=n+1}_{\rm dbl}+  \sum_{k=1}^{n-1} \tilde{\Phi}^{\rm m=n}_{\rm hole}, \,\, \mathrm{for} \,\, n<M_{\rm depth}   \\   
     \Phi(\Omega^n) &=& \sum_{k=1}^{n-1} \tilde{\Phi}^{\rm m=n}_{\rm hole} + \Phi^{\rm m=n}, \,\, \mathrm{for} \,\, n=M_{\rm depth}.
     \label{App:last}
\end{eqnarray}
Figure~\ref{fig:App0} presents the corresponding mesh decomposition for the case of three nested grids. The solution procedure is now split into six steps. Three steps shown in the top row of the figure are similar to those of the original scheme (see Fig~\ref{fig:5}). We calculate the potential $\Phi^{m=3}$ on the finest $\Omega^3$ grid, the potential $\Phi^{\rm m=1}_{\rm hole}$ on the coarsest $\Omega^1_{\rm hole}$ grid with a central hole, and the potential $\Phi^{\rm m=2}_{\rm hole}$ on the intermediate $\Omega^2_{\rm hole}$ grid with a central hole. The other three steps shown in the bottom row of Figure~\ref{fig:App0} involve the calculation of the potentials on the auxiliary grids, which are introduced for the correct calculation of the dipole moments.  
We note that the doubled $\Omega^2_{\rm dbl,hole}$ grid has now a central hole, which is filled with zero densities to avoid counting twice the mass distribution on the finest $\Omega^3$ grid. The doubled $\Omega^3_{\rm dbl}$ grid is analogous to $\Omega^2_{\rm dbl}$ grid for the case of two nested meshes. Finally, the $\Omega^3_{\rm qdp}$ grid is introduced, which is an extension of the finest $\Omega^3$ grid to the physical size of the coarsest $\Omega^1$ grid. The additional cells are filled with zero densities. This grid allows us to correctly calculate the dipole moments at the coarsest $\Omega^1$ grid. When the dipole moments are calculated across three neighboring grids, the runtime increases by a factor of 4.5 for $N=64$ and by a factor of 8.5 for $N=128$. An extension of the scheme to a larger number of neighboring grids can be done by analogy to Eqs.~(\ref{App:first})--(\ref{App:last}).

We note that the calculation of $\Phi^{m=3}_{\rm qdp}$ comes at a substantial computational expense in both the computational time and the memory requirements. 
A solution to this problem may be in switching to direct summation when calculating the input of the mass distribution on the finest $\Omega^3$ grid to the total gravitational potential on the coarsest $\Omega^1$ grid. In this case, the calculation of the potential $\Phi^3_{\rm qdp}$ on the quadrupled grid $\Omega^3_{\rm qdp}$ becomes obsolete. Instead, the input of the finest $\Omega^3$ grid to the potential on the coarsest $\Omega^1$ grid is calculated as
\begin{eqnarray}
    \Phi_{\rm d.s.}^{3\rightarrow 1}(x_i,y_j,z_k) &=& - G \sum_{i^\prime,j^\prime,k^\prime} M^{m=3}(x_{i^\prime},y_{j^\prime},z_{k^\prime})  \nonumber \\
    &\times& {\cal G}^{3\rightarrow1}(x_i-x_{i^\prime},y_j-y_{j^\prime},z_k-z_{k^\prime} )  ,
\end{eqnarray}
where $M^{m=3}$ is the mass distribution on the finest $\Omega^3$ grid and ${\cal G}^{3\rightarrow1}$ is the inverse distance between the cells on the $\Omega^1$ grid with coordinates $(x_i,y_j,z_k)$ and the cells on the $\Omega^3$ grid with coordinates ($x_i^\prime,y_j^\prime,z_k^\prime$). Our numerical experiments reveal that for a numerical resolution of $N\le 32^3$ the direct summation is faster than the calculation of $\Phi^3_{\rm qdp}$ using the convolution theorem.

\section{Further comparison of the modified and original convolution methods}
\label{App:compare}
\begin{figure}
\centering
\includegraphics[width=1\columnwidth]{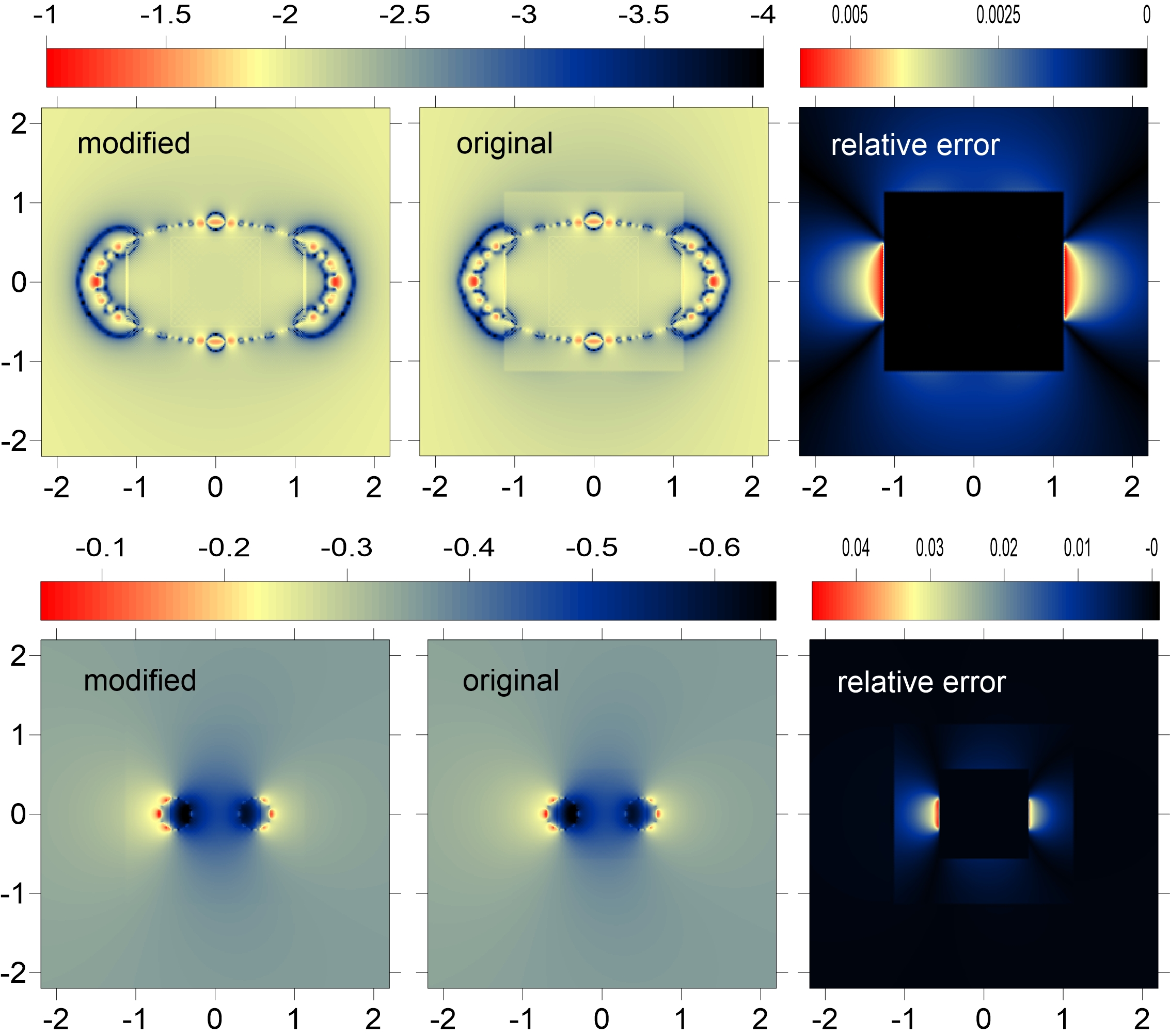}
\caption{Comparison of the modified and original convolution method for an oblate ellipsoid (top row) and wide separation binary (bottom row). The errors in the gravitational potential compared to the analytical solution are shown in the left and middle columns (in log units), while the right column displays the relative error in the gravitational potential between the two methods in percent. 
}
\label{fig:App4}
\end{figure}

Here, we perform an additional comparison between the original and modified convolution methods, the latter considering the dipole moments at every pair of neighboring grids (see Eqs.~\ref{App:first}--\ref{App:first2}). Figure~\ref{fig:App4} presents the 
errors in the gravitational potential relative to analytic solutions for each method and also the relative errors in the gravitational potential between the two methods. Four nested grids with a numerical resolution of $N=128^3$ for each grid were used in this test problem. The mass distributions similar to those considered in Sect.~\ref{Sect:performance} were used: an oblate ellipsoid and a wide separation binary. In particular, the ratio of semi-major axes of the ellipsoid was increased to 0.75:1.5:1.5 compared to the previous test problem (0.5:1.0:1.0) so the ellipsoid now covers all grid interfaces. 

In the modified method, a smoother transition between the grid interfaces can be noted, but otherwise the potentials are similar. The relative error between the two methods for a smooth density distribution (ellipsoid, top row) is very small, $<0.006\%$, indicating that the modification to the convolution method in this case is not really required. For a wide separation binary (bottom row), the error increased somewhat but still stays quite low, $<0.045\%$. Therefore, the modified method is recommended for models where the formation of tightly spaced gravitating objects is expected, for example, close binary stars or gravitationally fragmenting protostellar disks.

\end{appendix}

\end{document}